\documentclass[useAMS,usenatbib,fleqn]{mn2e}
\usepackage{epsfig}
\usepackage{deluxetable} 
\usepackage{url}
\usepackage{rotating}
\usepackage{amsmath}
\usepackage{graphicx}
\usepackage{setspace}
\usepackage{psfig}
\usepackage{natbib}
\usepackage{float}
\usepackage{placeins}
\usepackage{booktabs}

\title[Short time-scale variability of NLSy1 with UVOT]{Short time-scale variability of $\gamma$-ray-emitting narrow-line Seyfert 1 galaxies in optical and UV bands}
\author[D'Ammando]{F. D'Ammando$^{1}$\thanks{E-mail: dammando@ira.inaf.it} \\
$^{1}$INAF -- Istituto di Radioastronomia, Via Gobetti 101, I-40129 Bologna, Italy\\
}

\begin{document}

\date{Accepted. Received; in original form}

\maketitle

\label{firstpage}

\begin{abstract}

We report the first systematic analysis of single exposures of all optical and ultraviolet (UV) observations performed by the UltraViolet and Optical Telescope (UVOT) on board the {\em Neil Gehrels Swift Observatory} satellite available up to 2019 April of six $\gamma$-ray-emitting narrow-line Seyfert 1 galaxies (NLSy1). Rapid variability has been significantly detected on hours time-scale for 1H\,0323$+$342, SBS\,0846$+$513, PMN\,J0948$+$0022, and PKS\,2004--447 in 18 observations for a total of 34 events. In particular, we report the first detection of significant variability on short time-scale (3--6 ks) in optical for PKS\,2004--447, and UV for 1H\,0323$+$342 and PMN\,J0948$+$0022. The shortest variability time-scale observed for 1H\,0323$+$342, SBS\,0846$+$513, PMN\,J0948$+$0022, and PKS\,2004--447 (assuming a Doppler factor $\delta$ = 10) gives a lower limit on the size of emission region between 9.7$\times$10$^{14}$ (for SBS\,0846$+$513) and 1.6$\times$10$^{15}$ cm (for 1H\,0323$+$342), suggesting that the optical and UV emission during these events is produced in compact regions within the jet. These observations provide unambiguous evidence about the relativistically beamed synchrotron emission in these sources, similar to blazars. A remarkable variability has been observed for PMN\,J0948$+$0022 on 2009 June 23 with an increase from $\sim$1.1 to 0.4 mag going from $v$ to $w2$ filter in $\sim$1.6 h and a decrease at the initial level in a comparable time. The higher fractional flux change observed for this and other events at lower frequencies suggests that the synchrotron emission is more contaminated by thermal emission from accretion disc at higher frequencies. 
  
\end{abstract} 


\section{Introduction}

Narrow-line Seyfert 1 galaxies (NLSy1) are a class of active galactic nuclei (AGN) characterized by peculiar optical characteristics, that is, full width at half-maximum of the H$\beta$ broad emission line $<$ 2000 km s$^{-1}$, the flux ratio of [O III] to H$\beta$ $<$ 3, and a strong Fe$_{II}$ bump \citep[see e.g.][for a review]{pogge00}. Only a small fraction of NLSy1 are radio loud \citep[$<$ 7$\%$,][]{komossa06, zhou06, rakshit17}, a fraction lower than that estimated for quasars and broad-line AGN \citep[15--20 per cent, e.g.,][]{jiang07, kellermann16}. Observations with the Large Area Telescope (LAT) on board the {\em Fermi Gamma-ray Space Telescope} have revealed NLSy1 as a new class of $\gamma$-ray-emitting AGN with blazar-like properties   \citep{abdo09,dammando12,dammando15a, ajello20}. It is a very small class of objects up to now. Nine NLSy1\footnote{New deep spectroscopic observations of TXS 2116$-$077 disfavour its classification as an NLSy1 \citep{jarvela20}.} are included in the Fourth {\em Fermi} LAT source catalogue \citep[4FGL;][]{abdollahi20}, even considering the Data Release 2 of the 4FGL catalogue\footnote{https://fermi.gsfc.nasa.gov/ssc/data/access/lat/10yr\_catalog/}, which covers 10 yr (i.e., 2008 August 4--2018 August 4) of LAT data. Past studies have suggested that NLSy1 are powered by high-accretion processes \citep[e.g.,][]{peterson00,grupe04}, therefore investigating the properties of these sources is of interest for understanding similarities and differences with the other $\gamma$-ray-emitting AGN. Compared to the population of blazars, which includes BL Lacs and flat spectrum radio quasars (FSRQ), NLSy1 seems to share properties similar to FSRQ, but with typically lower $\gamma$-ray luminosities \citep[e.g.,][]{foschini15,dammando16,dammando19}.  
 
\begin{table*}
\caption{The number of {\em Swift}--UVOT observations with multiple exposures in the same epoch in each UVOT filter for the six sources studied, and total number of events studied. Redshift of each source and central wavelength of each filter are reported.}
\label{UVOT_Obs}
\begin{center}
\begin{tabular}{lcccccccc}
\hline
\multicolumn{1}{l}{\textbf{Source name}} &
\multicolumn{1}{c}{\textbf{$z$}}    &
\multicolumn{1}{c}{\textbf{$v$  (5468 \AA)}}  &
\multicolumn{1}{c}{\textbf{$b$  (4392 \AA)}}  &
\multicolumn{1}{c}{\textbf{$u$  (3465 \AA)}}  &
\multicolumn{1}{c}{\textbf{$w1$ (2600 \AA)}}  &
\multicolumn{1}{c}{\textbf{$m2$ (2246 \AA)}}  &
\multicolumn{1}{c}{\textbf{$w2$ (1928 \AA)}}  &
\multicolumn{1}{c}{\textbf{Total events}} \\
\hline                                      
1H\,0323$+$342      & $0.061$ & 86 & 76 & 80 & 79 & 70 & 97 & 488 \\ 
SBS\,0846$+$513     & $0.584$ & 20 & 22 & 23 & 27 & 22 & 22 & 136 \\ 
PMN\,J0948$+$0022   & $0.585$ & 23 & 23 & 42 & 24 & 23 & 23 & 158 \\
PKS\,1502$+$036     & $0.408$ &  3 &  3 &  6 & 6  & 4  &  8 & 30 \\
FBQS\,J1644$+$2619  & $0.145$ &  4 &  4 &  6 & 6  & 5  &  5 & 30 \\
PKS\,2004--447      & $0.240$ &  5 &  5 & 11 & 11 & 13 & 10 & 55 \\ 
\hline 
\end{tabular}
\end{center}
\end{table*} 

In \citet{dammando20}, we have analysed the optical, ultraviolet (UV), and X-ray observations performed by the {\em Neil Gehrels Swift Observatory} during 2006 July--2019 April of six $\gamma$-ray-emitting NLSy1 (i.e., 1H\,0323$+$342, SBS\,0846$+$513, PMN\,J0948$+$0022, PKS\,1502$+$036, FBQS\,J1644$+$2619, and PKS\,2004--447) to investigate their flux variability and spectral changes. A significant flux change has been observed in the optical and UV bands on time-scale of one day for 1H\,0323$+$342 and a few days for SBS\,0846$+$513, PMN\,J0948$+$0022, and PKS\,1502$+$036, clearly related to an increase of the non-thermal emission from a relativistic jet. On a long-term scale, a large variability amplitude has been observed for all six $\gamma$-ray-emitting NLSy1, significantly higher than the variability observed in other radio-loud NLSy1 not detected in $\gamma$-rays. These results have confirmed the dominance of the jet emission in the optical and UV part of their spectrum. 

Intraday optical variability has been investigated in the past and detected in 1H\,0323$+$342 \citep{itoh14,ojha19}, SBS\,0846$+$513 \citep{maune14,paliya16}, and PMN\,J0948$+$0022 \citep{liu10,itoh13,maune13}, with variations higher than 0.3 mag within a few hours, while no optical intraday variability has been detected in PKS\,1502$+$036 \citep{ojha19}. Variability amplitude of few per cent (4--16 per cent) on time-scales of 3--5 h has been reported by \citet{ojha20} for 1H\,0323$+$342, PMN\,J0948$+$0022, and FBQS\,J1644$+$2619.
                                                                                                                                                                                                                                                                 
Here, we present a systematic analysis on short time-scale of all optical and UV observations performed by the UltraViolet and Optical Telescope \citep[UVOT;][]{roming05} on board {\em Swift} for the six $\gamma$-ray-emitting NLSy1 studied in \citet{dammando20} in the same period (i.e. 2006 July--2019 April). The paper is organized as it follows. Section~\ref{Swift_obs} describes the {\em Swift}--UVOT observations and data analysis. Optical and UV variability results are reported and discussed in Section~\ref{variability}, while in Section~\ref{summary} we summarize our results. 
 
\section{{\em Swift}--UVOT observations and data analysis}\label{Swift_obs}

The {\em Neil Gehrels Swift Observatory} satellite \citep{gehrels04} observed the $\gamma$-ray-emitting NLSy1 1H\,0323$+$342, SBS\,0846$+$513, PMN\,J0948$+$0022, PKS\,1502$+$036, FBQS\,J1644$+$2619, and PKS\,2004--447 for 136, 32, 45, 16, 11, and 32 epochs, respectively, up to 2019 April. The results of the analysis of X-Ray Telescope observations have been presented in \citet{dammando20}. In this work, we focus on UVOT observations. In particular, we analyze the data corresponding to observations with multiple exposures with the same filter at the same epoch searching for variability on subdaily time-scales. Table~\ref{UVOT_Obs} reports the number of observations with multiple exposures of the six sources studied here separated for each filter, and the total number considering all filters together. In Appendix~\ref{UVOT_Appendix}, we report the results of the analysis of the images obtained by summing the multiple exposures with the same filter at the same epoch.
 
During the {\em Swift} pointings, the UVOT instrument observed the sources in its optical ($v$, $b$, and $u$) and UV ($w1$, $m2$, and $w2$) photometric bands \citep{poole08,breeveld10}. UVOT data in all filters were analysed with the \texttt{uvotmaghist} task included in the {\sc HEASoft} package (v6.26.1) and the 20170922 CALDB-UVOTA release. Source counts were extracted from a circular region of 5 arcsec radius centred on the source, while background counts were derived from a circular region with 20 arcsec radius in a nearby source-free region. All UVOT exposures were checked for possible small scale sensitivity (`sss') problems, which occur when the source falls on small detector regions where the sensitivity is lower\footnote{https://swift.gsfc.nasa.gov/analysis/uvot\_digest/sss\_check.html}. This issue has been observed in 28 of the 286 epochs analysed and only in the UV filters, as expected because the bad areas cover $\sim$7 per cent of the central 5 $\times$ 5 arcmin square of the field of view in UV filters, while $\sim$0.7 per cent of the central area in optical filters is affected. Exposures affected by the `sss' issue have been excluded in the following analysis.

After having analysed each single exposure, we have searched for significant ($>$ 3$\sigma$) changes of magnitude between two consecutive exposures collected with the same filter at the same epoch. The significance is calculated as $\Delta$mag/$\sqrt{(\sigma_{\rm\,1}^{2}+\sigma_{\rm\,2}^{2})}$, where $\Delta$mag = $|$ mag$_{\rm\,1}$ -- mag$_{\rm\,2}$ $|$, mag$_{\rm\,1}$ and mag$_{\rm\,2}$ are the magnitude of the two exposures, and $\sigma_{\rm\,1}$ and $\sigma_{\rm\,2}$ are the corresponding uncertainties. Systematic errors due to uncertainties in the photometric zero points are included in the total uncertainties considered. In Table~\ref{rapid}, we report for each event the date and filter, magnitude, and significance of detection for the single exposure, $\Delta$mag, significance of change between the two exposures, and the time elapsed between the two exposures. 
   
\begin{figure*}
\begin{center}
{\includegraphics[width=0.70\textwidth]{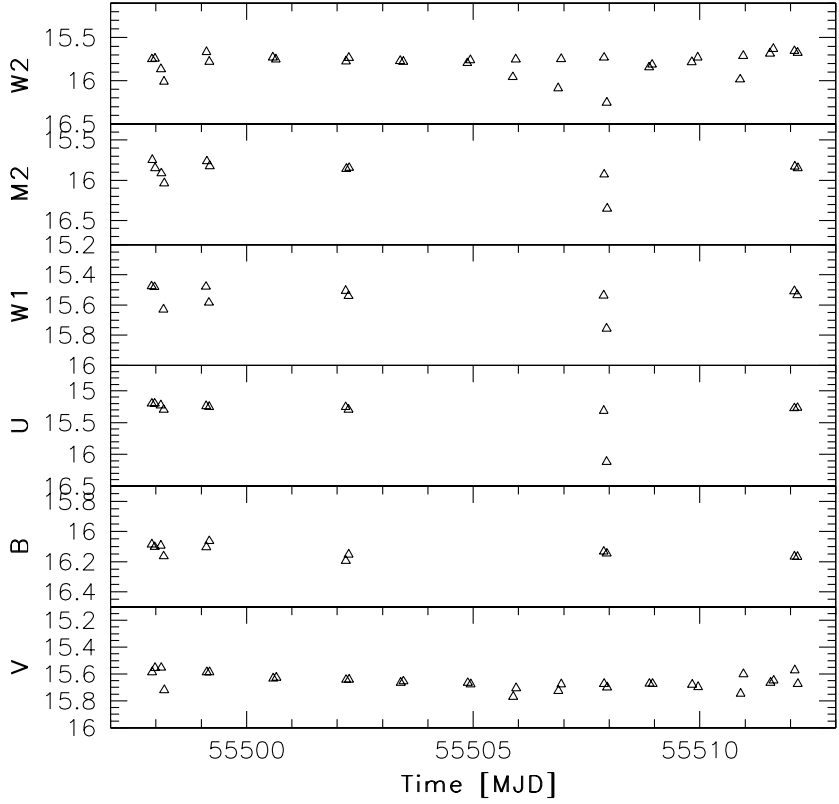}}
\caption{Light curve of 1H\,0323$+$342 in the six UVOT filters for the period 2010 October 28 (MJD 55497)--November 13 (MJD 55513) considering single exposures. Errors are small (0.04--0.06 mag), therefore are not shown in the plot.}
\label{0323_lc_zoom}
\end{center}
\end{figure*}

\setcounter{table}{1}
\begin{table*}
\caption{Results of the search for significant change of magnitude in consecutive UVOT exposures in the same epoch with the same filter for the six $\gamma$-ray-emitting NLSy1.}
\label{rapid} 
\begin{center}
\begin{tabular}{cccccccccc}
\hline
\textbf{Source name}   & \multicolumn{2}{c}{\textbf{Date}} & \multicolumn{3}{c}{\textbf{Single exposures}} & \multicolumn{3}{c}{\textbf{Change of magnitude}} \\
                       &  Gregorian  & MJD &  Filter    & Mag &  Significance$_{\rm\,image}$ ($\sigma$) &  $\Delta$mag & Significance$_{\rm\,change}$ ($\sigma$) & $\Delta$T (s) & \\
\hline
1H\,0323$+$342    &  2010-11-05  &  55505   &  $w2$   & 15.958 $\pm$ 0.043 &  48.9 &         &      &       \\
                  &              &          &         & 15.753 $\pm$ 0.042 &  56.2 &  0.205  & 3.4  & 5786  \\ 
                  &  2010-11-06  &  55506   &  $w2$   & 16.085 $\pm$ 0.043 &  47.8 &         &      &       \\
                  &              &          &         & 15.748 $\pm$ 0.042 &  56.3 &  0.337  & 5.6  & 5700  \\ 
                  &  2010-11-07  &  55507   &  $u$    & 15.312 $\pm$ 0.042 &  40.9 &         &      &       \\ 
                  &              &          &         & 16.118 $\pm$ 0.054 &  25.5 &  0.804  & 11.8 & 5798  \\ 
                  &              &          &  $w1$   & 15.537 $\pm$ 0.049 &  34.5 &         &      &       \\
                  &              &          &         & 15.755 $\pm$ 0.053 &  29.5 &  0.218  &  3.0 & 5812  \\ 
                  &              &          &  $m2$   & 15.929 $\pm$ 0.056 &  26.4 &         &      &       \\
                  &              &          &         & 16.352 $\pm$ 0.065 &  20.4 &  0.423  & 4.9  & 5716  \\ 
                  &              &          &  $w2$   & 15.732 $\pm$ 0.045 &  43.2 &         &      &       \\                              
                  &              &          &         & 16.252 $\pm$ 0.050 &  32.0 &  0.520  & 7.7  & 5762  \\ 
                  &  2010-11-10  &  55510   &  $w2$   & 15.984 $\pm$ 0.043 &  51.2 &         &      &       \\ 
                  &              &          &         & 15.710 $\pm$ 0.042 &  53.9 &  0.274  & 4.6  & 5973  \\ 
                  &  2010-11-28  &  55528   &  $w2$   & 15.930 $\pm$ 0.047 &  37.6 &         &      &       \\ 
                  &              &          &         & 15.554 $\pm$ 0.043 &  48.4 &  0.376  & 5.9  & 17345 \\ 
                  &  2013-03-02  &  56353   &  $b$    & 16.205 $\pm$ 0.047 &  31.2 &         &      &       \\                 
                  &              &          &         & 15.974 $\pm$ 0.044 &  34.0 &  0.231  & 3.5  & 5778  \\ 
                  &              &          &  $u$    & 15.437 $\pm$ 0.046 &  33.7 &         &      &       \\ 
                  &              &          &         & 15.244 $\pm$ 0.045 &  33.6 &  0.193  & 3.0  & 5789  \\ 
                  &              &          &  $w2$   & 15.971 $\pm$ 0.048 &  34.8 &         &      &       \\ 
                  &              &          &         & 15.764 $\pm$ 0.049 &  32.3 &  0.207  & 3.0  & 5713  \\ 
                  &  2015-09-17  &  57282   &  $u$    & 15.409 $\pm$ 0.053 &  25.2 &         &      &       \\ 
                  &              &          &         & 15.187 $\pm$ 0.050 &  29.2 &  0.222  & 3.0  & 7154  \\ 
                  &  2015-11-26  &  57352   &  $u$    & 15.221 $\pm$ 0.042 &  40.4 &         &      &       \\ 
                  &              &          &         & 15.454 $\pm$ 0.054 &  26.1 &  0.234  & 3.4  & 6568  \\ 
SBS\,0846$+$513   &  2013-04-22  &  56404   &  $v$    & 16.388 $\pm$ 0.049 &  26.2 &         &      &       \\
                  &              &          &         & 15.972 $\pm$ 0.040 &  36.5 &  0.416  & 6.6  & 5143  \\  
                  &              &          &  $u$    & 16.517 $\pm$ 0.053 &  25.7 &         &      &       \\
                  &              &          &         & 16.265 $\pm$ 0.054 &  25.2 &  0.252  & 3.3  & 5118  \\
PMN\,J0948$+$0022 &  2009-05-05  &  54956   &  $u$    & 16.542 $\pm$ 0.073 &  16.9 &         &      &       \\ 
                  &              &          &         & 16.914 $\pm$ 0.091 &  12.9 &  0.372  & 3.2  & 5768  \\  
                  &  2009-06-14  &  54996   &  $u$    & 17.212 $\pm$ 0.136 &   8.3 &         &      &       \\  
                  &              &          &         & 16.660 $\pm$ 0.099 &  11.8 &  0.552  & 3.3  & 5760  \\
                  &  2009-06-23  &  55005   &  $v$    & 18.174 $\pm$ 0.262 &   4.2 &         &      &       \\   
                  &              &          &         & 17.051 $\pm$ 0.122 &   9.2 &  1.123  & 3.9  & 5807  \\  
                  &              &          &  $b$    & 18.303 $\pm$ 0.132 &   8.5 &         &      &       \\ 
                  &              &          &         & 17.469 $\pm$ 0.081 &  14.6 &  0.834  & 5.4  & 5907  \\ 
                  &              &          &  $u$    & 17.545 $\pm$ 0.103 &  11.2 &         &      &       \\
                  &              &          &         & 16.878 $\pm$ 0.076 &  16.0 &  0.667  & 5.2  & 5928  \\
                  &              &          &  $w1$   & 17.474 $\pm$ 0.090 &  13.2 &         &      &       \\ 
                  &              &          &         & 16.770 $\pm$ 0.070 &  18.2 &  0.704  & 6.6  & 5979  \\
                  &              &          &  $m2$   & 17.367 $\pm$ 0.089 &  13.5 &         &      &       \\
                  &              &          &         & 16.844 $\pm$ 0.076 &  16.5 &  0.523  & 4.5  & 5778  \\
                  &              &          &  $w2$   & 17.366 $\pm$ 0.064 &  21.0 &         &      &       \\
                  &              &          &         & 16.947 $\pm$ 0.059 &  23.8 &  0.419  & 4.8  & 5857  \\                
\hline
\end{tabular}                  
\end{center}
\end{table*}                  
                  
\addtocounter{table}{-1}                  
\begin{table*}
\caption{Results of the search for significant change of magnitude in consecutive UVOT exposures in the same epoch with the same filter for the six $\gamma$-ray-emitting NLSy1. Note: The asterisk $^{*}$ indicates that multiple exposures are present only for the filters reported in table.}
\begin{center}
\begin{tabular}{cccccccccc}
\hline
\textbf{Source name}   & \multicolumn{2}{c}{\textbf{Date}} & \multicolumn{3}{c}{\textbf{Single exposures}} & \multicolumn{3}{c}{\textbf{Change of magnitude}} \\
                       &  Gregorian  & MJD &  Filter    & Mag &   Significance$_{\rm\,image}$ ($\sigma$) &  $\Delta$mag & Significance$_{\rm\,change}$ ($\sigma$) & $\Delta$T (s)  \\
\hline
                                                                
PMN\,J0948$+$0022      &  2012-03-26  &  56012$^{*}$   &  $u$  & 17.778 $\pm$ 0.047 &  30.8 &        &      &        \\    
                       &              &                &       & 17.523 $\pm$ 0.043 &  37.3 & 0.255  & 4.0  & 17275  \\            
                       &  2012-03-30  &  56016$^{*}$   &  $u$  & 17.219 $\pm$ 0.039 &  46.3 &        &      &        \\                         
                       &              &                &       & 16.998 $\pm$ 0.037 &  52.4 & 0.221  & 4.1  & 5757   \\
                       &  2012-04-23  &  56040$^{*}$   &  $u$  & 17.404 $\pm$ 0.056 &  23.7 &        &      &        \\    
                       &              &                &       & 16.873 $\pm$ 0.068 &  18.4 & 0.531  & 6.0  & 30340  \\ 
                       &  2012-12-30  &  56291$^{*}$   &  $u$  & 16.196 $\pm$ 0.097 &  12.3 &        &      &        \\  
                       &              &                &       & 15.703 $\pm$ 0.067 &  19.1 & 0.494  & 4.2  &  5265  \\
                       &              &                &  $w1$ & 16.344 $\pm$ 0.057 &  25.3 &        &      &        \\
                       &              &                &       & 15.721 $\pm$ 0.072 &  17.8 & 0.623  & 6.8  &  5361  \\
                       &  2013-01-11  &  56303         &  $b$  & 18.175 $\pm$ 0.146 &   7.7 &        &      &        \\       
                       &              &                &       & 17.524 $\pm$ 0.074 &  16.3 & 0.651  & 4.0  & 46709  \\
                       &              &                &  $u$  & 17.411 $\pm$ 0.126 &   9.0 &        &      &        \\ 
                       &              &                &       & 16.672 $\pm$ 0.065 &  19.2 & 0.739  & 5.2  & 46658  \\       
                       &              &                &  $w1$ & 17.120 $\pm$ 0.108 &  10.8 &        &      &        \\ 
                       &              &                &       & 16.613 $\pm$ 0.068 &  19.3 & 0.507  & 4.0  & 46583  \\
                       &              &                &  $m2$ & 17.304 $\pm$ 0.113 &  10.2 &        &      &        \\
                       &              &                &       & 16.756 $\pm$ 0.075 &  16.8 & 0.548  & 4.0  & 47042  \\
                       &              &                &  $w2$ & 17.549 $\pm$ 0.097 &  12.2 &        &      &        \\
                       &              &                &       & 16.800 $\pm$ 0.058 &  24.4 & 0.749  & 6.6  & 46833  \\ 
PKS\,2004--447         &  2013-11-20  &  56616$^{*}$   &  $u$  & 18.310 $\pm$ 0.062 &  20.4 &        &      &        \\ 
                       &              &                &       & 18.665 $\pm$ 0.076 &  15.6 & 0.356  & 3.6  & 5820   \\                   
\hline                                               
\end{tabular}                                                            
\end{center}                                                         
\end{table*}                                                         

\section{Results and Discussion}\label{variability}

\begin{figure*}
\begin{center}
\rotatebox{0}{\resizebox{!}{57mm}{\includegraphics{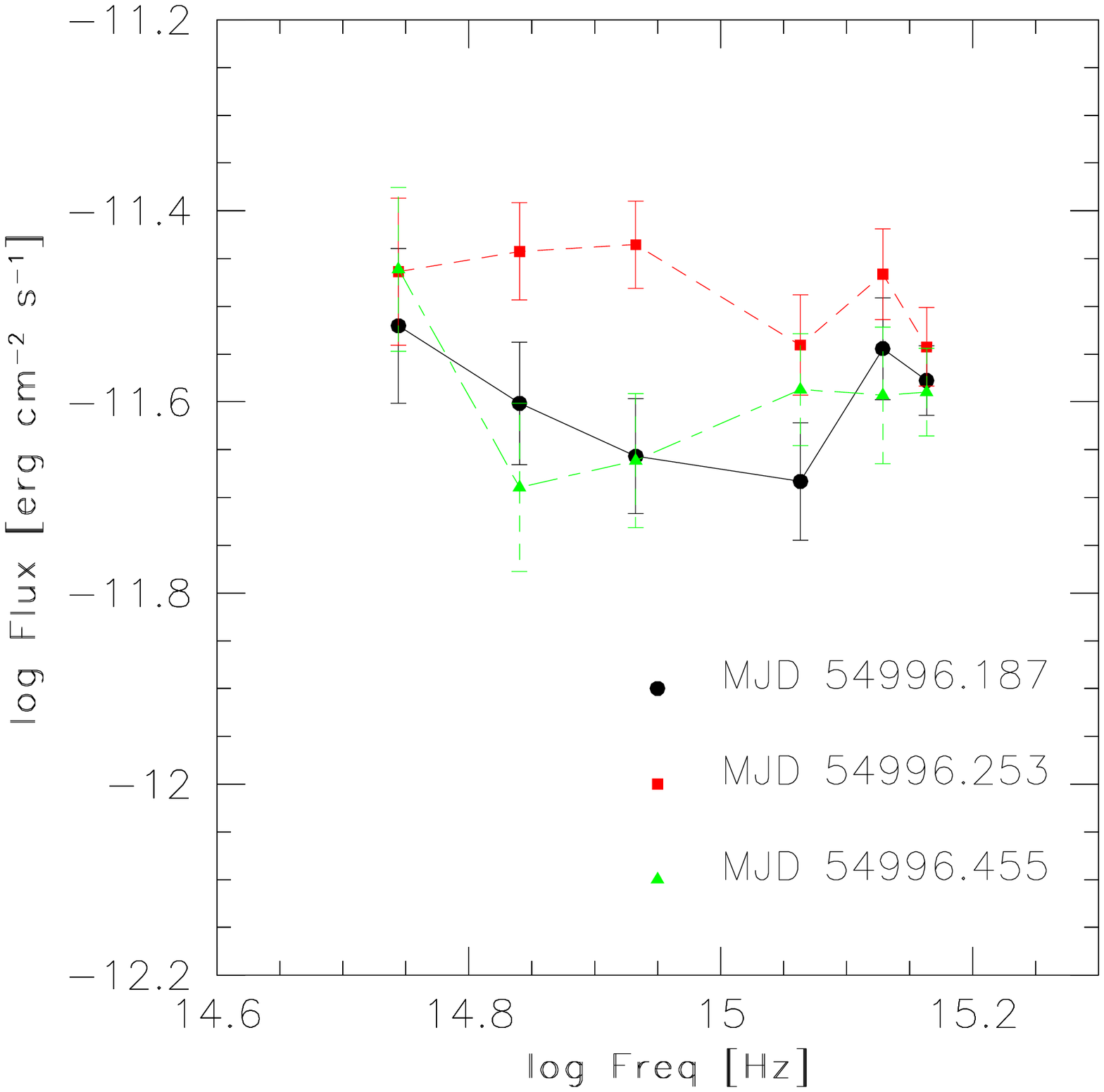}}}
\hspace{0.05cm}
\rotatebox{0}{\resizebox{!}{57mm}{\includegraphics{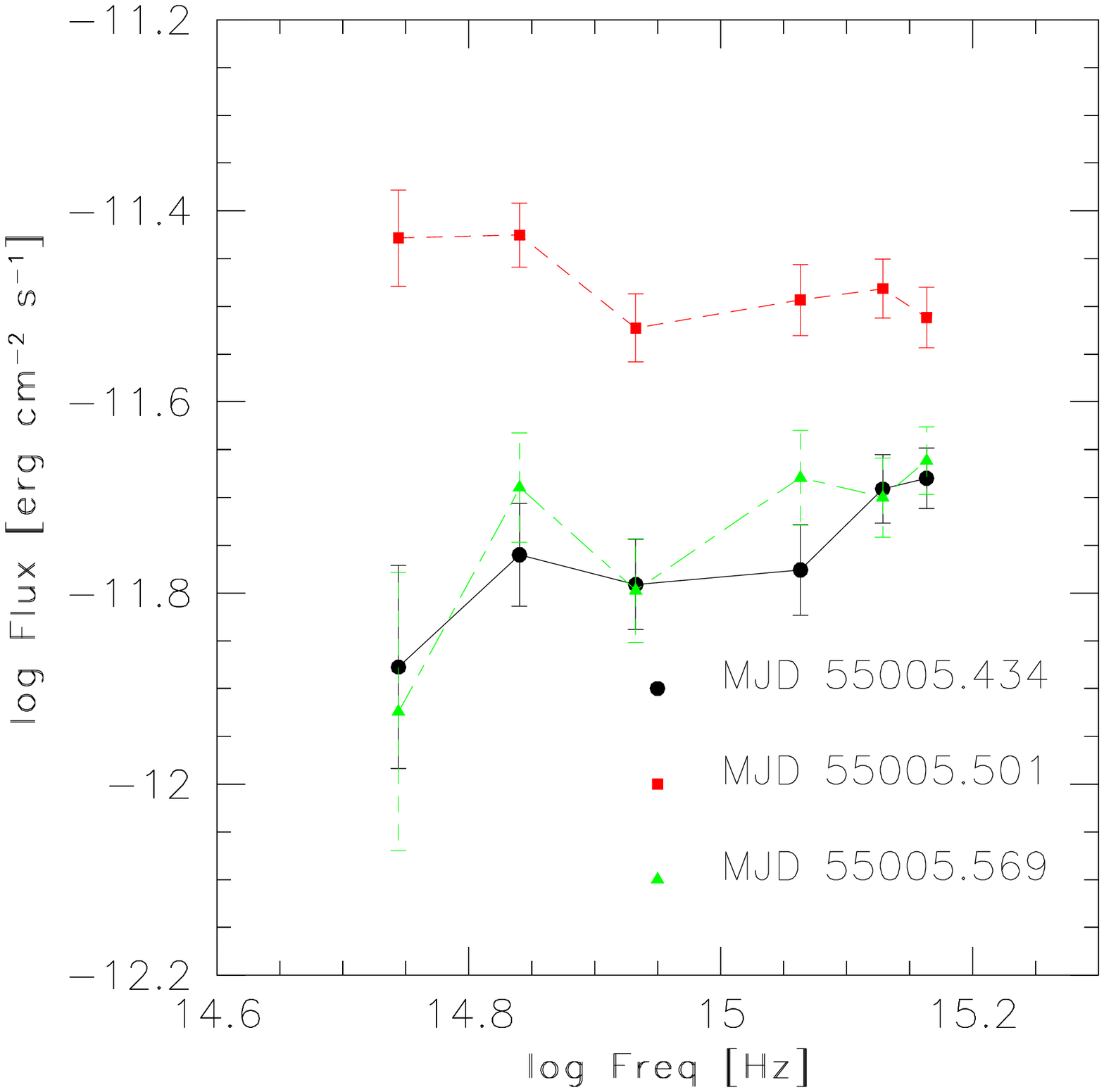}}}
\hspace{0.05cm}
\rotatebox{0}{\resizebox{!}{57mm}{\includegraphics{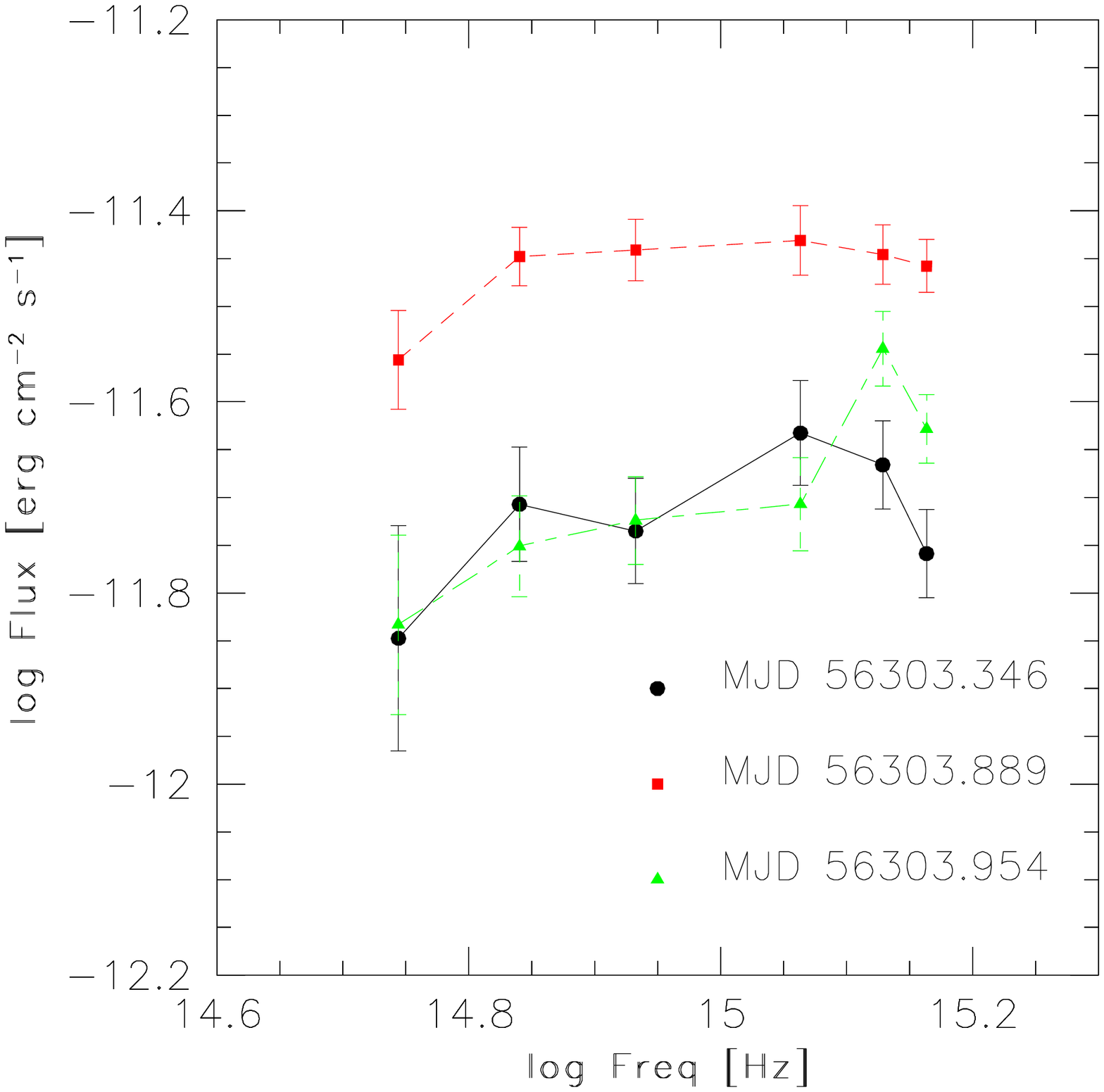}}}
\caption{SED of PMN\,J0948$+$0022 obtained by {\em Swift}--UVOT from $v$ to $w2$ filters on 2009 June 14 (MJD 54996; left-hand panel), 2009 June 23 (MJD 55005; centre panel), and 2013 January 11 (MJD 56003; right-hand panel). Fluxes are corrected for Galactic extinction. Different symbols and colours refer to images collected at different times within the same {\em Swift} observation.} 
\label{0948_SED}
\end{center}
\end{figure*}

\begin{table*}
\caption{Fractional change of flux, rest-frame time-scale, and short-term variability amplitude based on significant change of magnitude in consecutive UVOT exposures in the same epoch with the same filter for the six $\gamma$-ray-emitting NLSy1 reported in Table~\ref{rapid}. The short-term variability amplitude (V$_{\rm\,amp,\,ST}$) is calculated as the ratio between the peak magnitude observed in each epoch and the average magnitude estimated over the entire period (see Table~\ref{Amplitude_summed}).}                                            
\label{rapid2} 
\begin{center}
\begin{tabular}{ccccccc}
\hline
\textbf{Source name}   & \multicolumn{2}{c}{\textbf{Date}} &  \textbf{Filter} & \textbf{Fractional Change} & \textbf{$\Delta$T$_{\rm\,rest}$} & \textbf{$V_{\rm\,amp,\,ST}$} \\
                  &  Gregorian   &  MJD     &         & $\%$    & (s)  & \\
\hline
1H\,0323$+$342    &  2010-11-05  &  55505   &  $w2$   &  21   &  5453  &  1.03  \\
                  &  2010-11-06  &  55506   &  $w2$   &  36   &  5372  &  1.03  \\
                  &  2010-11-07  &  55507   &  $u$    &  110  &  5465  &  1.03  \\ 
                  &              &          &  $w1$   &  22   &  5478  &  1.02  \\
                  &              &          &  $m2$   &  48   &  5387  &  0.96  \\ 
                  &              &          &  $w2$   &  61   &  5431  &  1.05  \\                             
                  &  2010-11-10  &  55510   &  $w2$   &  29   &  5630  &  1.07  \\ 
                  &  2010-11-28  &  55528   &  $w2$   &  41   &  16348 &  1.24  \\ 
                  &  2013-03-02  &  56353   &  $b$    &  24   &  5445  &  1.23  \\                 
                  &              &          &  $u$    &  19   &  5456  &  1.10  \\ 
                  &              &          &  $w2$   &  21   &  5385  &  1.02  \\ 
                  &  2015-09-17  &  57282   &  $u$    &  23   &  6743  &  1.16  \\ 
                  &  2015-11-26  &  57352   &  $u$    &  24   &  6190  &  1.12  \\ 
SBS\,0846$+$513   &  2013-04-22  &  56404   &  $v$    &  46   &  3247  &  11.89 \\
                  &              &          &  $u$    &  26   &  3231  &  14.29 \\ 
PMN\,J0948$+$0022 &  2009-05-05  &  54956   &  $u$    &  41   &  3639  &  2.09  \\ 
                  &  2009-06-14  &  54996   &  $u$    &  66   &  3634  &  1.87  \\  
                  &  2009-06-23  &  55005   &  $v$    &  181  &  3664  &  1.96  \\   
                  &              &          &  $b$    &  116  &  3727  &  1.85  \\ 
                  &              &          &  $u$    &  85   &  3740  &  1.52  \\
                  &              &          &  $w1$   &  91   &  3772  &  1.41  \\ 
                  &              &          &  $m2$   &  62   &  3645  &  1.40  \\
                  &              &          &  $w2$   &  47   &  3695  &  1.28  \\
                  &  2012-03-26  &  56012   &  $u$    &  26   &  10899 &  0.82  \\    
                  &  2012-03-30  &  56016   &  $u$    &  23   &  3632  &  1.36  \\        
                  &  2012-04-23  &  56040   &  $u$    &  63   &  19142 &  1.53  \\    
                  &  2012-12-30  &  56291   &  $u$    &  58   &  3322  &  4.63  \\  
                  &              &          &  $w1$   &  78   &  3382  &  3.80  \\
                  &  2013-01-11  &  56303   &  $b$    &  82   &  29469 &  1.76  \\       
                  &              &          &  $u$    &  98   &  29437 &  1.85  \\ 
                  &              &          &  $w1$   &  60   &  29390 &  1.63  \\ 
                  &              &          &  $m2$   &  66   &  29679 &  1.52  \\
                  &              &          &  $w2$   &  99   &  29548 &  1.48  \\
PKS\,2004--447    &  2013-11-20  &  56616   &  $u$    &  39   &  4127  &  1.87  \\                 
\hline                                                         
\end{tabular}                                                                  
\end{center}                                                              
\end{table*}                                                              

\begin{figure*}
\begin{center}
\rotatebox{0}{\resizebox{!}{57mm}{\includegraphics{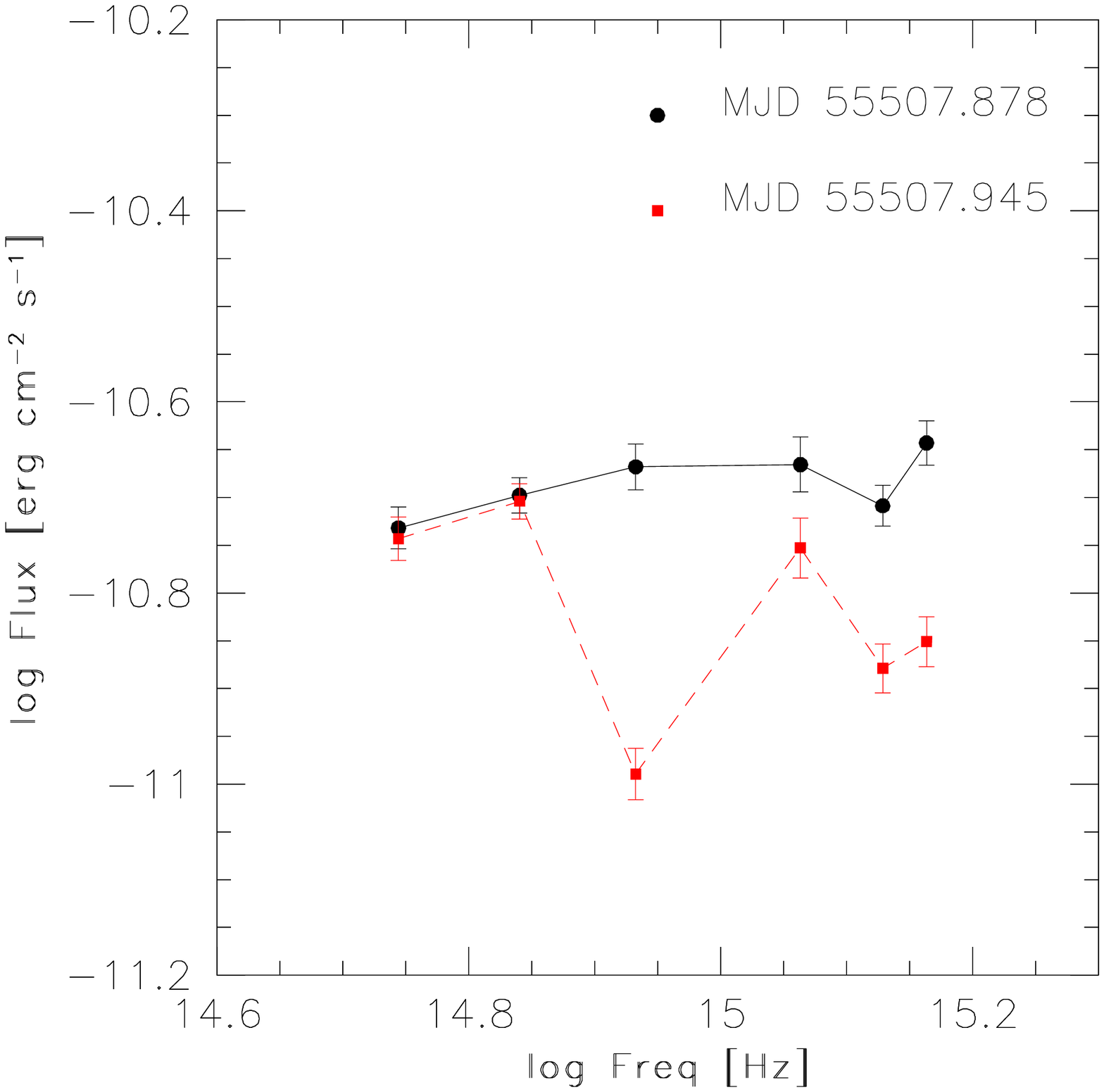}}}
\hspace{0.05cm}
\rotatebox{0}{\resizebox{!}{57mm}{\includegraphics{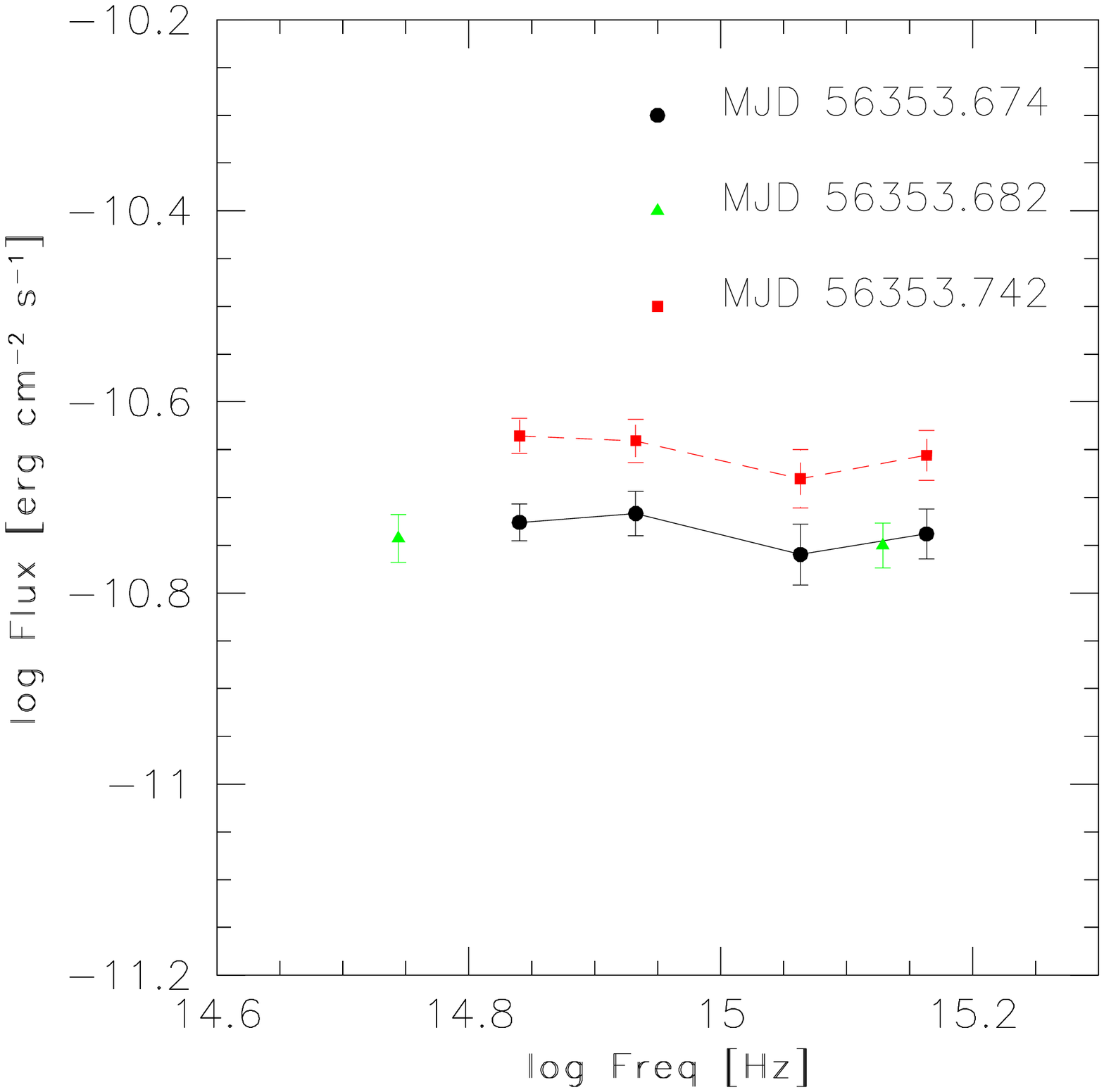}}}
\hspace{0.05cm}
\rotatebox{0}{\resizebox{!}{57mm}{\includegraphics{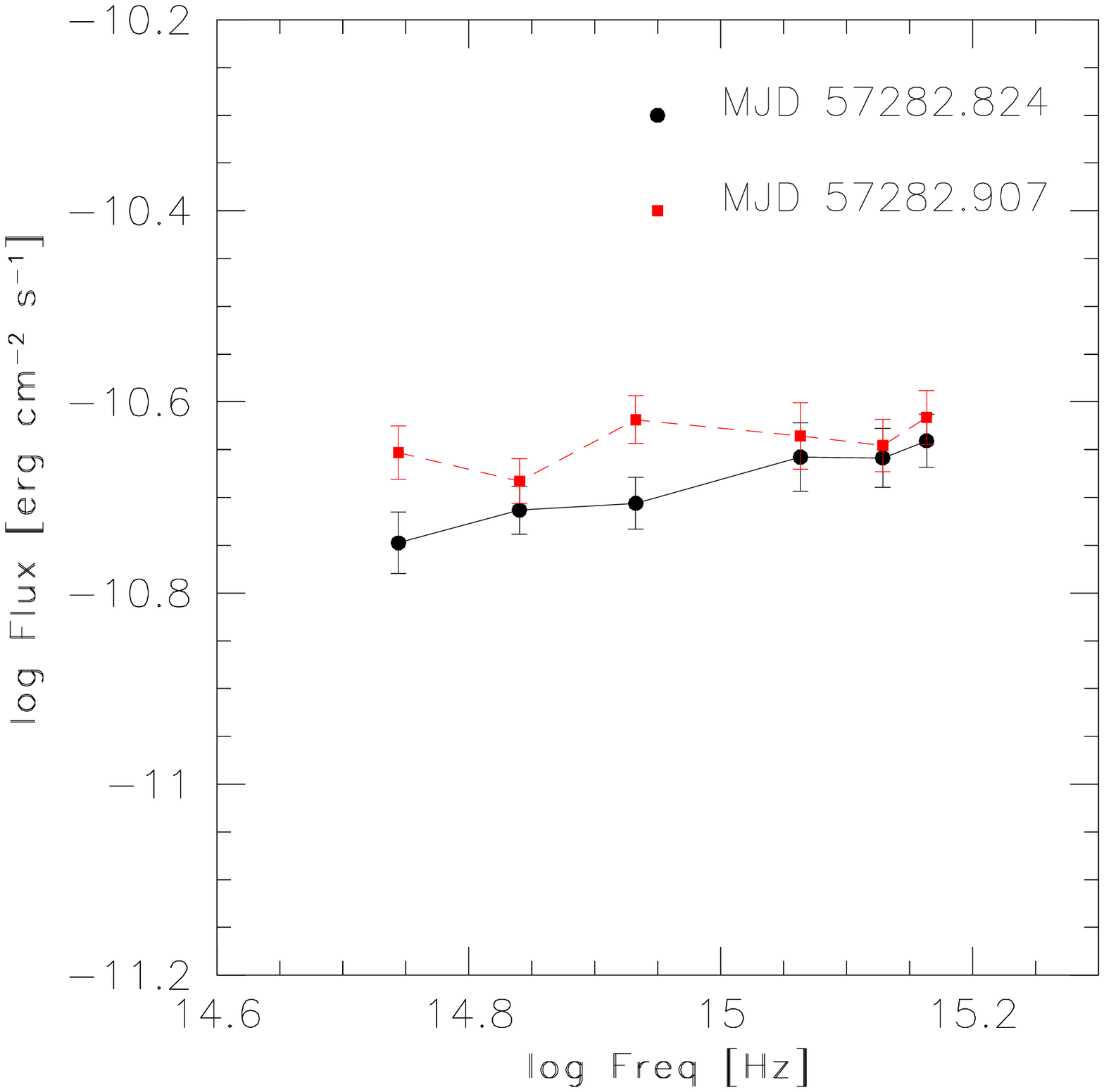}}}
\caption{SED of 1H\,0323$+$323 obtained by {\em Swift}--UVOT from $v$ to $w2$ filters on 2009 June 25 (MJD 55507; left-hand panel) and 2013 March 2 (MJD 56353; centre panel), and 2015 July 17 (MJD 57282; right-hand panel). Fluxes are correct for Galactic extinction. Different symbols and colours refer to images collected at different times within the same {\em Swift} observation.}
\label{0323_SED}
\end{center}
\end{figure*}

\begin{figure*}
\begin{center}
{\includegraphics[width=0.75\textwidth]{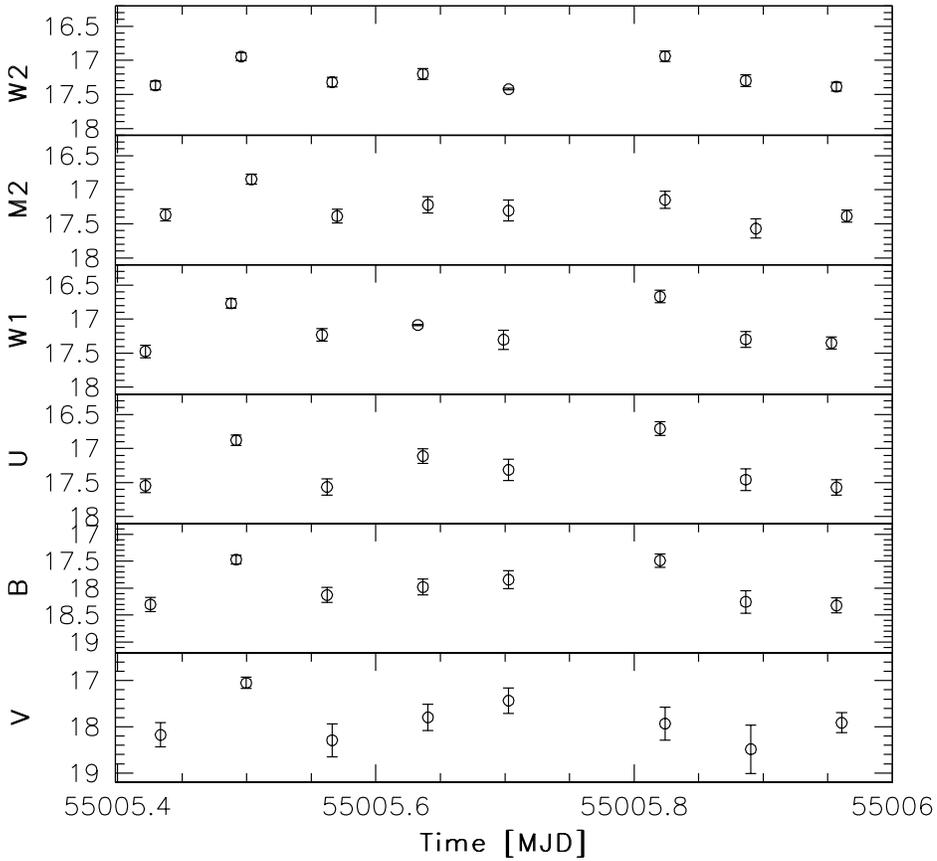}}
\caption{Light curve of the single exposures of PMN\,J0948$+$0022 in the six UVOT filters collected on 2009 June 23 (MJD 55005).}
\label{0948_event1}
\end{center}
\end{figure*}

\begin{figure}
\begin{center}
{\includegraphics[width=0.5\textwidth]{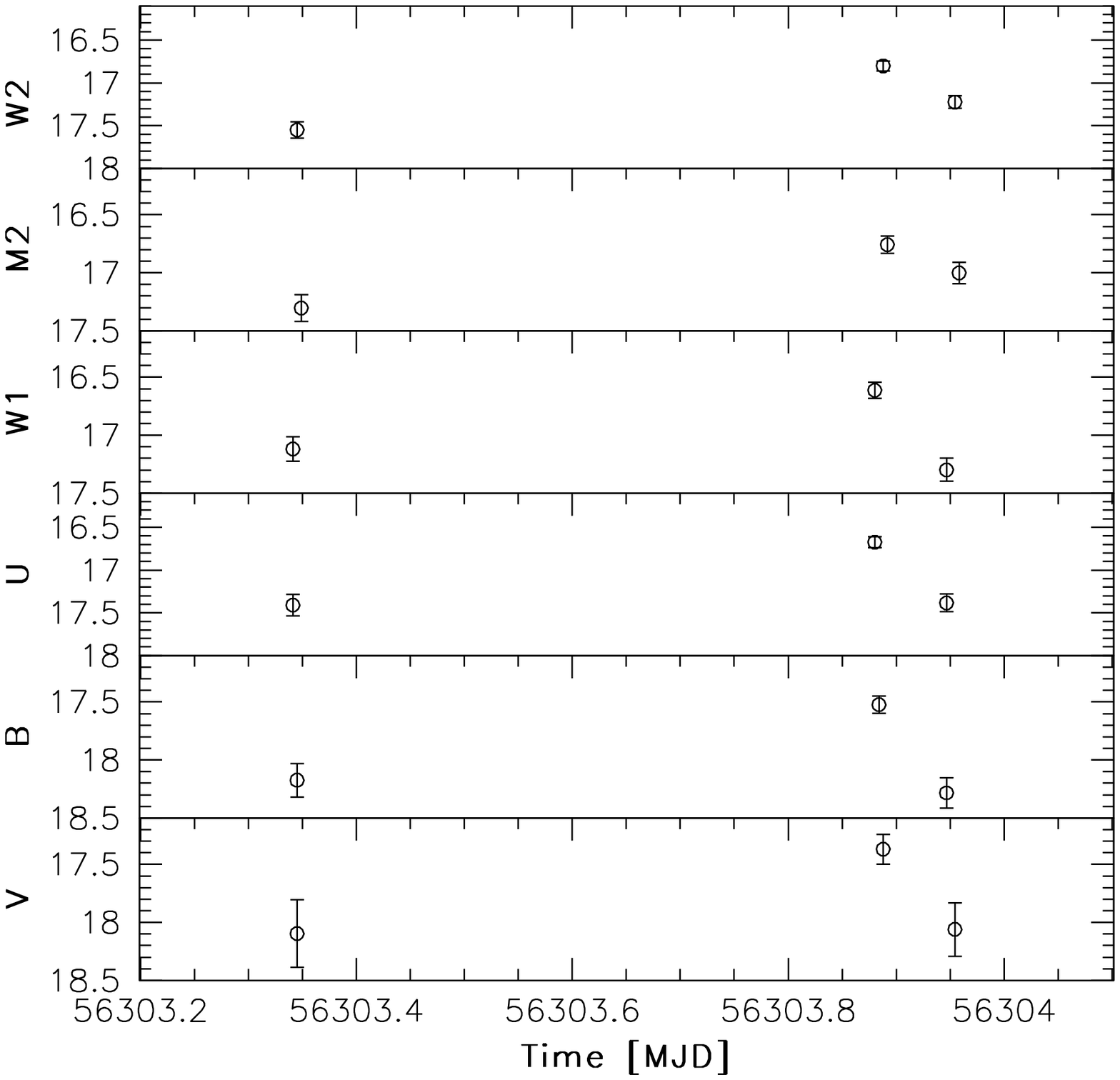}}
\caption{Light curve of the single exposures of PMN\,J0948$+$0022 in the six UVOT filters collected on 2013 January 11 (MJD 56303).}
\label{0948_event2}
\end{center}
\end{figure}

Significant variability has been observed on hours time-scale for 1H\,0323$+$342, SBS\,0846$+$513, PMN\,J0948$+$0022, and PKS\,2004--447 in 18 observations for a total of 34 events taken into account each filter separately. In particular, a change of magnitude has been detected in eight observations for 1H\,0323$+$342 (for a total of 13 events considering the different filters), in one observation for SBS\,0846$+$513 (for a total of 2 events considering the different filters), in eight observations for PMN\,J0948$+$0022 (for a total of 18 events considering the different filters), and in one observation for PKS\,2004--447 (for a total of 1 event considering the different filters). In 4 of the 18 observations (on 2010 November 10 and 2015 November 26 for 1H\,0323$+$342; on 2009 May 5 for PMN\,J0948$+$0022; on 2013 November 20 for PKS\,2004--447) we have seen a decrease of the activity between the two exposures on a time-scale of $\sim$6 ks. In those cases, the lack of observations before the tentative peak of activity leaves open the possibility of a more rapid rising time with respect to the decreasing part of the activity observed by {\em Swift}. In case of FBQS\,J1644$+$2619, a change of 0.474 mag has been observed on 2015 April 9 in the $m2$ filter, but at a significance of 2.2$\sigma$. No changes at a significance $>$ 2$\sigma$ has been observed for PKS\,1502$+$036. 

Careful checks made for each single image ensure that these events are not related to instrumental artifacts. Three factors can influence the detection of such a rapid change of magnitude with UVOT: the number of observations with multiple images collected at the same epoch, the average brightness of the source, and its activity state in the period studied. 

The number of observations with multiple exposures in optical and UV filters is comparable (442 observations summing $v$, $b$, and $u$ filters, 455 observations summing $w1$, $m2$, and $w2$ filters), although due to the `sss' issue some events in the UV filters have been discarded. This reflects in a comparable number of events in optical (19 events, 14 of them observed in $u$) and UV (15 events, 8 of them observed in $w2$) filters. Considering each object separately, we notice that for PKS\,1502$+$036, FBQS\,J1644$+$2644, and PKS\,2004--447, the number of observations with multiple exposures is relatively low (30, 30, and 55, respectively; see Table~\ref{UVOT_Obs}), while it is significantly higher for 1H 0323$+$342, SBS\,0846$+$513, and PMN\,J0948$+$0022 (488, 136, and 158, respectively; see Table~\ref{UVOT_Obs}). However, the percentage of events detected for the latter three sources is quite different: $\sim$3 per cent, $\sim$1.5 per cent, and $\sim$12 per cent, respectively. This suggests that the average brightness of the source, and thus the uncertainties obtained, can play an important role in revealing variability on short-term time-scales. Considering the analysis of the images obtained by summing the single exposures collected at each epoch (see Table A1--A6 in Appendix~\ref{UVOT_Appendix}) the median of the uncertainties are: 0.06 mag for 1H\,0323$+$342, 0.08 mag for PMN\,J0948$+$0022, 0.10 mag for FBQS\,J1644$+$2619, 0.19 mag for PKS\,1502$+$036, 0.22 mag for PKS\,2004--447, and 0.24 mag for SBS\,0846$+$513. Uncertainties are even larger when considering the single images instead of the summed images. This can explain the low number of events observed for SBS\,0846$+$513. In this context, we would have expected a larger number of events for 1H 0323$+$342. However, the smaller percentage of events, if compared to PMN J0948$+$0022, may be due to the very low long-term variability amplitude observed for this source over the entire {\em Swift} monitoring (see Table~\ref{Amplitude_summed} in Appendix~\ref{UVOT_Appendix}). 

In Table~\ref{rapid2}, we report the short-term variability amplitude of these rapid events calculated as the ratio between the peak magnitude observed in each epoch and the average magnitude estimated over the entire period. Differently from what observed for SBS\,0846$+$513, PMN\,J0948$+$0022, and PKS\,2004--447, in case of 1H\,0323$+$342 the short-term variability amplitude is quite low for each event. It is worth mentioning that only for two events (on MJD 55507 in $m2$ filter for 1H\,0323$+$342, and on MJD 56012 in $u$ filter for  PMN\,J0948$+$0022) the source has been observed with a lower magnitude with respect to average value obtained over the entire {\em Swift} monitoring, confirming that almost all the short time-scale variability events are detected during high activity states.
                                                                   
The magnitude changes observed in the different filters (considering all the four sources in Table~\ref{rapid}) vary in the interval: 0.42--1.12 in $v$, 0.23--0.83 in $b$, 0.19--0.80 in $u$, 0.22--0.70 in $w1$, 0.42--0.55 in $m2$, and 0.21--0.75 in $w2$, with a higher amplitude of variability in the optical bands with respect to the UV ones. This is opposite to the behaviour usually observed in quasars, that is, increasing variability with increasing frequencies \citep[e.g.,][]{vandenberk04}, in agreement with a dominant contribution of the synchrotron emission in both optical and UV bands in these NLSy1. However, in this context, the larger changes observed in optical with respect to UV may be related also to the presence, in addition to variable emission from a jet, of an almost constant emission from an accretion disc at higher frequencies, as usually observed in FSRQ \citep[e.g.][]{dammando11,raiteri12}.  

For 1H\,0323$+$342 and PMN\,J0948$+$0022, changes of magnitude have been significantly detected in both optical and UV bands, at least one time for all six filters, except for the $v$ filter in case of 1H\,0323$+$342 (see Fig.~\ref{0323_lc_zoom}). On the other hand, for SBS\,0846$+$513 and PKS\,2004--447 significant variability has been observed only in optical bands. Considering that in the latter two sources a prominent accretion disc is missing \citep[see e.g.][]{dammando13,orienti15}, the lack of variability on hours time-scale in UV should not be related to the presence of thermal emission from accretion disc that can dilute the jet emission. For PKS\,1502$+$036 and FBQS\,J1644$+$2619, the lack of variability on hours time-scale can be mainly related to the small number of {\em Swift}--UVOT observations with multiple exposures  available at the same epoch. Moreover, in case of PKS\,1502$+$036, the source is relatively faint (see the average magnitude in Table~\ref{Average_summed} in Appendix~\ref{UVOT_Appendix}), therefore the large uncertainties obtained due to the short observation time of the single exposures prevent us to detect significant changes of magnitude. In case of FBQS\,J1644$+$2619, the long-term variability amplitude estimated over all {\em Swift} observations available is low ($<$ 2) in all filters (see Table~\ref{Amplitude_summed} in Appendix~\ref{UVOT_Appendix}), indicating a period of relatively low activity of the source during the {\em Swift} monitoring. 

In Table~\ref{rapid2}, we report the fractional flux change obtained for the 34 events identified in the UVOT data. 

For PMN\,J0948$+$0022, the fractional flux change is higher in the optical filters than in the UV ones. The spectral energy distribution (SED) of the source in optical and UV (from $v$ to $w2$ filter) is shown in Fig.~\ref{0948_SED} for three events with images in all filters, corresponding to MJD 54996 (left-hand panel), MJD 55005 (centre panel), and MJD 56303 (right-hand panel). The UVOT magnitudes are corrected for Galactic extinction using the $E(B-V)$ value from \citet{schlafly11} and the extinction laws from \citet{cardelli89} and converted to flux densities using the conversion factors from \citet{breeveld10}. The lower fractional flux change in UV can be related to the presence of a bright accretion disc with a luminosity of L$_{\rm\,disc}$= 5.7$\times$10$^{45}$ erg s$^{-1}$ \citep{dammando15b} peaking in the UV part of the spectrum that can partly dilute the jet emission reducing the variation of the emission. The presence of an accretion disc may be seen in the UV part of the three SED obtained before the short-term variability events (MJD 54996.187, 55005.434, 56303.346). A possible shift of the synchrotron peak to higher frequencies is observed at the time of the short-term variability events on MJD 54996.253 and MJD 56303.889. After the variability event, the optical and UV flux levels are comparable to those before the event. On MJD 56303.954, a hint of the accretion disc seems to be present 1.6 h after the peak flux.

For 1H\,0323$+$342, the behaviour observed during the short-term variability events may be consistent with jet emission that dominates in the optical and a significant contribution from the accretion disc in the UV. However, a more complicated behaviour seems to emerge in some events. A higher fractional flux change has been observed on MJD 55507 in the $u$ filter with respect to the UV band, but no evidence of variability is detected in the $v$ and $b$ filters (Fig.~\ref{0323_SED}, left-hand panel). Owing to the low redshift of the source ($z$ = 0.061), the host galaxy may contribute to the total emission in optical bands. Therefore, the thermal emission from the host may dilute the variability amplitude in the $v$ and partly $b$ filters \citep[see e.g.,][]{ojha20}. On MJD 57282, a change of magnitude has been observed only in the optical band (in particular in $u$ filter and at lower significance in the $v$ filter; Fig~\ref{0323_SED}, right-hand panel). On the contrary a comparable fractional flux change has been observed in optical ($b$ and $u$ filters) and UV ($w1$ and $m2$ filters) bands on MJD 56353 (Fig~\ref{0323_SED}, centre panel)\footnote{On MJD 56353 only one image has been collected in the $v$ and $w2$ filters.}. This chromatic behaviour is likely intrinsic, and is related to injection/acceleration of particles in the jet. No significant change of the overall shape of the optical--UV spectrum has been observed between the two exposures on MJD 57282 and MJD 56353.

Comparing PMN\,J0948$+$0022 and 1H\,0323$+$342, a higher fractional flux change has been observed in the former (an average fractional flux change of 75 per cent) with respect to the latter (an average fractional flux change of 37 per cent), in agreement with the higher long-term variability amplitude observed in the different filters over the long-term monitoring provided by {\em Swift} (see Table~\ref{Amplitude_summed} in Appendix~\ref{UVOT_Appendix}). Differently from the other $\gamma$-ray-emitting NLSy1, over a long time-scale the variability amplitude of 1H\,0323$+$342 increases with increasing frequencies, and this may be related to a less powerful jet when the source is not in a high activity period, and a larger contribution of a quite bright accretion disc \citep[1.4 $\times$ 10$^{45}$ erg s$^{-1}$,][]{abdo09}. Moreover, as previously reported, a steady thermal emission from the host galaxy may contribute to the optical emission with a  dilution of the optical synchrotron emission.
                                                                                                                                                                                                                                                                                                                                                                                                                                                                                                                                                                                                                                                                                                                                                                                                                                                                                                          
\noindent Notes on individual sources are reported below.
  
{\it 1H\,0323$+$342}: the source has been monitored on daily basis between 2010 October 28 and November 30 (see Fig.~\ref{0323_lc} in Appendix~\ref{UVOT_Appendix}), showing five variability events in $w2$, the best sampled filter, and in particular four events between 2010 November 5 and 10 (see Fig.~\ref{0323_lc_zoom}). No significant events have been observed in $v$ filter, where the number of exposures is comparable to the $w2$ filter. The lower variability observed in $v$ filter (and more generally in the optical part of the spectrum, see the long-term variability amplitude in Table ~\ref{Amplitude_summed} in Appendix ~\ref{UVOT_Appendix}) for 1H\,0323$+$342 can be due to a higher contamination from thermal emission from the host galaxy. Moreover, in this source the accretion disc can contribute more significantly to the total emission in optical--UV. In the other filters only one event has been observed in 2010 November, but it can be due to a sparse sampling.

{\it SBS\,0846$+$513}: notwithstanding the source has shown the largest long-term variability amplitude in the optical and UV bands during the {\em Swift} monitoring (see Table~\ref{Amplitude_summed} in Appendix~\ref{UVOT_Appendix}), a change of magnitude has been significantly detected only in one observation. This is mainly due to the relatively large uncertainties on magnitude obtained for the single exposures related also to the relatively faintness of the source (see the average magnitudes in Table~\ref{Average_summed} of Appendix~\ref{UVOT_Appendix}). Moreover, in case of 2013 April 22, the exposures in $w1$ and $w2$ filters have been affected by the `sss' issue not allowing us to properly compare the variability in optical and UV bands. A few other events with a large change of magnitude but at a significance of 2 $<$ $\sigma$ $<$ 3 have been observed, with a $\Delta$mag of 0.606 (on 2013 January 30) and 0.974 (on 2019 March 20) in $u$ filter, 0.893 (on 2019 January 16), 0.635 (on 2011 August 30), and 0.584 (on 2011 January 30) in $m2$ filter, 0.576 (on 2019 March 6) and 0.641 (on 2019 April 3) in $w2$ filter.

{\it PMN\,J0948$+$0022}: a spectacular variability has been observed on 2009 June 23 (see Fig.~\ref{0948_event1}) with an increase from $\sim$1.1 to 0.4 mag going from $v$ to $w2$ filter in $\sim$1.6 h (corresponding to 1 h in rest frame, see Table~\ref{rapid2}) and a decrease at the initial level in a comparable time (see also Fig.~\ref{0948_SED}, centre panel). The approximate symmetry of the event, with a similar rising and decaying time, suggests that the relevant time-scale in this case can be the light crossing time of the emitting region \citep[e.g.][]{chiaberge99}. A second peak with a smaller amplitude and at lower significance has been observed a few hours later. A similar rapid variability in optical bands has been reported for PMN\,J0948$+$0022 in \citet{liu10} and \citet{eggen13}. The lower variability amplitude observed at higher frequencies implies that the synchrotron emission is more contaminated by thermal emission from accretion disc at higher frequency, as expected by the SED modelling of the source \citep[see e.g.][]{dammando15b}. Another flaring event from this source on a longer time-scale is shown in Fig.~\ref{0948_event2} (left-hand panel), with a decreasing variability going to higher frequencies. 
In addition, several other events at a significance of 2 $<$ $\sigma$ $<$ 3 have been observed. In particular, on 2009 May 5 simultaneously to a significant change in $u$ band, a $\Delta$mag of 0.430 and 0.427 in $v$ and $b$ filters has been observed with a significance of 2.5$\sigma$ and 2.9$\sigma$, respectively. In the same way, on 2009 June 14, simultaneously to a significant change in $u$ band, there is a $\Delta$mag of 0.724, 0.385, and 0.256 in $b$, $w1$, and $w2$ filters with a significance of 2.9$\sigma$, 2.5$\sigma$, and 2.1$\sigma$, respectively (see also Fig.~\ref{0948_SED}, left-hand panel). Moreover, it is worth mentioning that on 2009 June 4, a $\Delta$mag of 0.869, 0.437, and 0.440 has been observed in $v$, $b$, and $w1$ filters with a significance of 2.4$\sigma$, 2.5$\sigma$, and 2.3$\sigma$, respectively. The higher number of events observed in $u$ filter is related to the fact that 21 of the 45 {\em Swift}--UVOT observations are performed only in that filter.

{\it PKS\,2004--447}: similarly to SBS\,0846$+$513, the quite large uncertainties on the magnitudes estimated in single exposures, related to the relatively faintness of the source (see Table~\ref{Average_summed} in Appendix~\ref{UVOT_Appendix}) reduce the significance of the change of magnitude observed. A few other events with a large change of magnitude but at a significance of 2 $<$ $\sigma$ $<$ 3 have been observed, with a $\Delta$mag of 0.387 (on 2012 September 30) and 0.791 (on 2014 March 14) in $w1$ filter, and 0.712 (on 2013 October 20) and 0.854 (on 2016 May 1) in $w2$ filter.

Large amplitude variability events on short time-scales in radio-quiet NLSy1 are usually rare \citep[e.g.,][]{klimek04}, although not completely absent in literature, as shown by optical variations of 0.1--0.2 mag of IRAS 13224--3809 on time-scales of an hour \citep{miller00}. In that case the rapid optical variability has been explained as due to hot spots, or flares, or magnetic reconnection in the accretion disc. No clear evidence of large amplitude short-term variability in UV band has been reported for radio-quiet NLSy1, with only small amplitude rapid variability observed for 1H\,0707--495 \citep{robertson15} and IRAS\,13224--3809 \citep{buisson18}. 

On the other hand, rapid optical variability has been observed in several blazars, the other class of $\gamma$-ray-emitting class of AGN with a relativistic jet pointed towards us \citep[e.g.,][]{miller89, carini90, heidt96, romero02, sagar04}. Such behaviour is usually observed during a flaring activity, but it may also be an observational bias due to the fact that intensive monitoring of blazars in optical is carried out mainly when the activity of the source is high. Similar events in UV are less common for blazars \citep[e.g.,][]{edelson92}, although this can  be related to the lack of dedicated monitoring programs in that energy range. 

\noindent The detection of rapid variability in optical in blazars has been explained by invoking a shock moving down the jet \citep[e.g.,][]{marscher85}, turbulent cells behind a shock in the jet \citep[e.g.,][]{marscher14}, irregularities in the jet flow such as mini-jets \citep[e.g.,][]{giannios09}, or variations in the outflow parameters due to magnetic reconnection \citep[e.g.,][]{sironi15, christie19}. In all these scenarios it is important the Doppler boosting of the jet emission, which amplifies the amplitude and shorten the time-scale of the variation, and therefore the viewing angle of the emitting region has to be close to the observer's line of sight \citep[see e.g.][]{raiteri17}. In this context, the rapid and large variability episodes detected for these $\gamma$-ray-emitting NLSy1 clearly point to the presence of a relativistically beamed jet, closely aligned to our line of sight, similarly to blazars.

The detection of polarized emission has been considered the incontrovertible proof that the synchrotron emission dominates in the optical part of the spectrum of blazars \citep[e.g.,][]{itoh16, zhang19}. Polarization as high as 10 per cent and 36 per cent has been detected in the $\gamma$-ray-emitting NLSy1 SBS\,0846$+$513 \citep{maune14} and PMN\,J0948$+$0022 \citep{itoh13} accompanied by intraday variation in the photometric light curves. For 1H\,0323$+$342, \citet{itoh14} reported an increase of the optical polarization degree from 0-1 per cent to 3 per cent during a high optical state. The lower polarization degree in this source may be due to the contamination from the thermal accretion disc emission also in the optical part of the spectrum. Significant polarization increase has been reported also for PKS\,1502$+$036 in \citet{angelakis18}. These results indicate the synchrotron origin from a compact region of the jet for the optical emission in these $\gamma$-ray-emitting NLSy1, at least during high activity states, confirming the similarity between these sources and blazars.  
 
Taking into account the cosmological redshift of the source, we calculate the intrinsic variability time-scale as $\Delta$T$_{\rm\,rest}$ = $\Delta$T / (1+$z$). This results in a minimum intrinsic variability time-scale of 3.2 ks for SBS\,0846$+$513, 3.3 ks for PMN\,J0948$+$0022, 4.1 ks for PKS\,2004--447, and 5.4 ks for 1H\,0323$+$342, as reported in Table \ref{rapid2}. 
Based on causality argument, it is possible to constrain the intrinsic size of the emitting region during these rapid variability events to be $R$ $<$ $c$\,$\delta$\,$\Delta$T$_{\rm\,rest}$) = 1.6--4.9 $\times$10$^{15}$\,cm for 1H\,0323$+$342, 9.7$\times$10$^{14}$\,cm for SBS\,0846$+$513, 1.1--8.8$\times$10$^{15}$\,cm for PMN\,J0948$+$0022, and 1.4$\times$10$^{15}$\,cm for PKS\,2004--447 (assuming a typical Doppler factor $\delta$ = 10). This suggests that the optical and UV emission during these events is produced in compact regions within the jet. It is worth mentioning that the orbital period of the {\em Swift} satellite is 5700 s, and usually one exposure per orbit in each filter has been collected with UVOT and during the same portion of the orbit, limiting the minimum time between two exposures collected with the same filter to 5--6 ks. Therefore, the observed variability time-scale estimated here may be an upper limit to the effective time-scale of the variation and the intrinsic size of the emitting region may be smaller. 

\noindent The minimum intrinsic variability estimated by the optical and UV observations for these NLSy1 can be also connected to the event horizon light crossing time of their supermassive black hole as t$_{\rm\,lc}$ $\sim$ r$_{\rm\,g}$/c $\sim$ GM$_9$/c$^{3}$ $\sim$ 1.4 $\times$ M$_9$ h, where r$_{\rm\,g}$ is the gravitational radius, M$_9$ = (M/10$^{9}$) M$_{\odot}$ is the BH mass in solar masses, and $c$ is the speed of light \citep[e.g.,][]{begelman08}. Assuming a BH mass of 10$^{9}$ M$_{\odot}$, t$_{\rm\,lc}$ $\sim$ 1.4 h. Comparing this value to the minimum intrinsic variability time-scale obtained for these sources provides an upper limit to their BH mass, which is a debated issue for $\gamma$-ray-emitting NLSy1 \citep[see][for a detailed discussion]{dammando19}. In particular, we obtain an upper limit of 1.1$\times$10$^{9}$ M$_{\odot}$ for 1H\,0323$+$342, 6.4$\times$10$^{8}$ M$_{\odot}$ for SBS\,0846$+$513, 6.6$\times$10$^{8}$ M$_{\odot}$ for PMN\,J0948$+$0022, and 8.2$\times$10$^{8}$ M$_{\odot}$ for PKS\,2004--447. A more dense short-term monitoring with more than one exposure per filter within the same orbit with {\em Swift} will be important to set tighter constraints to the minimum variability time-scale and therefore to the size of the emitting region and the BH mass of these objects. 

\section{Summary}\label{summary}

We have performed the first systematic analysis of single exposures of all optical and UV observations carried out by {\em Swift}--UVOT available up to 2019 April of six $\gamma$-ray-emitting NLSy1. Our main results are summarized below.

1. Significant ($>$ 3$\sigma$) magnitude changes have been observed on hours time-scale for 1H\,0323$+$342, SBS\,0846$+$513, PMN\,J0948$+$0022, and PKS\,2004--447 in 18 observations for a total of 34 events in different filters. After removing exposures affected by the `sss' issue, there are 205 UVOT observations with multiple exposures at least in one filter in the same epoch. Rapid variability has been observed in $\sim$9 per cent of the {\em Swift}--UVOT observations, with a percentage of events detected of $\sim$3 per cent, $\sim$1.5 per cent, and $\sim$12 per cent for 1H\,0323$+$342, SBS\,0846$+$513, and PMN J0948$+$0022, respectively. Several other magnitude changes at a significance level 2 $<$ $\sigma$ $<3$ have been observed. These events provide unambiguous evidence about the relativistically beamed synchrotron emission in these $\gamma$-ray-emitting NLSy1, similar to what is observed in blazars. During these events a higher amplitude of variability has been observed in the optical bands with respect to the UV ones, in agreement with a dominant contribution of the synchrotron emission.

2. We have detected for the first time rapid variability in optical on time-scale of $\sim$6 ks ($\sim$4.1 ks in the rest frame) in PKS\,2004--447, and for the first time rapid variability in UV on time-scale of $\sim$6 ks in 1H\,0323$+$342 and PMN\,J0948$+$0022, corresponding to 5.4 and 3.4 ks in the rest frame, respectively. 

3. For 1H\,0323$+$342 and PMN\,J0948$+$0022, short-term variability has been significantly detected in both optical and UV bands, at least one time for all six UVOT filters, except for the $v$ filter in 1H\,0323$+$342. A higher fractional flux change has been observed in PMN\,J0948$+$0022 with respect to 1H\,0323$+$342. For PMN\,J0948$+$0022, the fractional flux change is higher in the optical filters than in the UV ones during these rapid variability events. For 1H\,0323+342, the larger fractional flux changes have been observed in optical, where the jet emission should dominate, while a significant thermal emission from accretion disc is present in the UV part of the spectrum.
 
4. Notwithstanding SBS\,0846$+$513 has shown the largest long-term variability amplitude in the optical and UV bands over the entire {\em Swift} monitoring, a change of magnitude has been significantly detected only in one observation. This is mainly due to the fact that the source is quite faint, therefore the uncertainties for the single exposures are relatively large, reducing the significance of the variations. On the contrary, for the brightest object in optical and UV, 1H\,0323$+$342, although it did not show an extreme variability amplitude in the period studied, several episodes of short-term variability have been identified in both optical and UV bands.

5. For PKS\,1502$+$036 and FBQS\,J1644$+$2619 no significant variability has been detected on hours time-scale. This can be related to the small number of {\em Swift} observations available for these sources but also to the faintness of PKS\,1502$+$036 and the low activity of FBQS\,J1644$+$2619 in the period studied here.

6. Short variability time-scales can be used to extrapolate the size of the emitting region. The shortest variability time-scale observed in optical or UV for 1H\,0323$+$342, SBS\,0846$+$513, PMN\,J0948$+$0022, and PKS\,2004-447 (assuming a Doppler factor $\delta$ = 10) gives a lower limit on the size of emission region of R $<$ 1.6$\times$10$^{15}$, 9.7$\times$10$^{14}$, 1.1$\times$10$^{15}$, and 1.4$\times$10$^{15}$ \,cm, respectively. This is an indication that the optical and UV emission during these events is produced in compact regions within the jet.

7. A remarkable variability has been observed for PMN\,J0948$+$0022 on 2009 June 23 with an increase from 1.1 to 0.4 mag going from $v$ to $w2$ filter in $\sim$1.6 h and a decrease at the initial level in a comparable time. The higher variability amplitude observed at lower frequencies suggests that the synchrotron emission is more contaminated by thermal emission from accretion disc at higher frequencies. 

{\em Swift}--UVOT observations can be important for studying short-term variability in $\gamma$-ray-emitting NLSy1. From this first systematic analysis of six objects seems to emerge that an important factor for detecting rapid changes of magnitude is the average brightness of the target. In fact, the two sources for which the largest number of events have been detected are 1H \,0323$+$342 and PMN\,J0948$+$0022, the brightest sources in the sample. This is directly related to the uncertainties obtained for single images, with large uncertainties reducing the significance of the detection, even if the change of magnitude is large. Such a kind of studies can be applied also to blazars, the other class of jetted AGN seen at a small angle to our line of sight. This will allow us to compare the properties of these two classes of $\gamma$-ray-emitting AGN in terms of rapid variability observed in optical and UV bands.
 
\section*{Acknowledgements}

FD thanks the {\em Swift} team for making these observations possible, the duty scientists, and science planners. This research has made use of the XRT Data Analysis Software ({\sc XRTDAS}). This work made use of data supplied by the UK Swift Science Data Centre at the University of Leicester. This research has made use of data obtained through the High Energy Astrophysics Science Archive Research Center Online Service, provided by the NASA/Goddard Space Flight Center. This research has made use of the NASA/IPAC Extragalactic Database (NED), which is operated by the Jet Propulsion Laboratory, California Institute of Technology, under contract with the National Aeronautics and Space Administration. Part of this work is based on archival data, software or online services provided by the Space Science Data Center - ASI. FD thanks the anonymous referee for constructive comments and suggestions. 
 
\section*{Data Availability}

The data underlying this article are available in {\em Swift} Archive Download Portal at the UK {\em Swift} Science Data Centre (https://www.swift.ac.uk/swift\_portal/).

\appendix

\onecolumn
 
\section{Swift--UVOT summed images results}\label{UVOT_Appendix}

For each epoch, we processed the images obtained by summing the exposures with the same filter in the same epoch with the task \texttt{uvotimsum} and then aperture photometry was performed with the task \texttt{uvotsource}. All images are aspect corrected to ensure that individual exposures are summed without offsets\footnote{https://www.swift.ac.uk/analysis/uvot/image.php}. We extracted source counts from a circular region of 5 arcsec radius centred on the source, while background counts were derived from a circular region with 20 arcsec radius in a nearby source-free region. The observed magnitude obtained for the six sources are reported in Tables \ref{UVOT_0323}--\ref{UVOT_2004}. Values in italics refer to epochs in which multiple exposures with the same filter at the same epoch are available. 
 
\setcounter{table}{0}
\begin{table*}
\caption{Observed magnitude of 1H 0323$+$324 obtained by {\em Swift}--UVOT.}
\label{UVOT_0323}
\begin{center}
\begin{tabular}{cccccccc}
\hline
\multicolumn{1}{c}{\textbf{Date (UT)}} &
\multicolumn{1}{c}{\textbf{MJD}}       &
\multicolumn{1}{c}{\textbf{$v$}}       &
\multicolumn{1}{c}{\textbf{$b$}}       &
\multicolumn{1}{c}{\textbf{$u$}}       &
\multicolumn{1}{c}{\textbf{$w1$}}      &
\multicolumn{1}{c}{\textbf{$m2$}}      &
\multicolumn{1}{c}{\textbf{$w2$}}      \\
\hline
2006-07-06  &  53922  &  {\it 15.54 $\pm$ 0.04}  & {\it 16.11 $\pm$ 0.07}  & {\it 15.13 $\pm$ 0.05} & {\it 15.33 $\pm$ 0.05}  & {\it 15.63 $\pm$ 0.06}  & {\it 15.57 $\pm$ 0.07}   \\  
2006-07-09  &  53925  &            $-$           & {\it 16.04 $\pm$ 0.05}  & {\it 15.08 $\pm$ 0.05} & {\it 15.30 $\pm$ 0.05}  &          $-$            &           $-$            \\ 
2007-07-20  &  54301  &  {\it 15.66 $\pm$ 0.04}  & {\it 16.23 $\pm$ 0.05}  & {\it 15.36 $\pm$ 0.05} & {\it 15.61 $\pm$ 0.06}  & {\it 15.92 $\pm$ 0.06}  & {\it 15.82 $\pm$ 0.05}  \\       
2007-11-04  &  54410  &  {\it 15.72 $\pm$ 0.05}  & {\it 16.36 $\pm$ 0.05}  & {\it 15.58 $\pm$ 0.06} & {\it 15.86 $\pm$ 0.07}  & {\it 16.23 $\pm$ 0.08}  & {\it 16.09 $\pm$ 0.06}  \\  
2007-11-11  &  54417  &  {\it 15.82 $\pm$ 0.05}  & {\it 16.41 $\pm$ 0.05}  & {\it 15.56 $\pm$ 0.05} & {\it 15.79 $\pm$ 0.06}  & {\it 16.11 $\pm$ 0.08}  & {\it 15.98 $\pm$ 0.06}  \\ 
2007-11-25  &  54429  &  {\it 15.80 $\pm$ 0.05}  & {\it 16.34 $\pm$ 0.05}  & {\it 15.54 $\pm$ 0.05} & {\it 15.84 $\pm$ 0.06}  & {\it 16.21 $\pm$ 0.07}  & {\it 16.06 $\pm$ 0.06}  \\ 
2007-12-01  &  54435  &  {\it 15.81 $\pm$ 0.05}  & {\it 16.36 $\pm$ 0.05}  & {\it 15.53 $\pm$ 0.05} & {\it 15.84 $\pm$ 0.04}  & {\it 16.10 $\pm$ 0.07}  & {\it 16.01 $\pm$ 0.06}  \\
\hline
\end{tabular}
\end{center}
\end{table*} 

\addtocounter{table}{-1}
\begin{table*}
 \begin{center}
\begin{tabular}{cccccccc}
\hline
\multicolumn{1}{c}{\textbf{Date (UT)}} &
\multicolumn{1}{c}{\textbf{MJD}}       &
\multicolumn{1}{c}{\textbf{$v$}}       &
\multicolumn{1}{c}{\textbf{$b$}}       &
\multicolumn{1}{c}{\textbf{$u$}}       &
\multicolumn{1}{c}{\textbf{$w1$}}      &
\multicolumn{1}{c}{\textbf{$m2$}}      &
\multicolumn{1}{c}{\textbf{$w2$}}      \\
\hline 
2012-01-30  &  55956  &  15.65 $\pm$ 0.07        & {\it 16.26 $\pm$ 0.05}   & {\it 15.38 $\pm$ 0.05}  & {\it 15.66 $\pm$ 0.07}  & 15.92 $\pm$ 0.09        & {\it 15.97 $\pm$ 0.07}  \\
2012-03-03  &  55989  &  15.85 $\pm$ 0.06        &  16.48 $\pm$ 0.06        &  15.65 $\pm$ 0.06       & 15.90 $\pm$ 0.07        & 16.30 $\pm$ 0.08        & 16.22 $\pm$ 0.07        \\ 
2013-01-13  &  56305  &  {\it 15.77 $\pm$ 0.05}  &  {\it 16.33 $\pm$ 0.05}  &  {\it 15.50 $\pm$ 0.05} & {\it 15.75 $\pm$ 0.06}  & {\it 16.13 $\pm$ 0.07}  & {\it 15.99 $\pm$ 0.06}  \\
2013-01-14  &  56306  &  {\it 15.69 $\pm$ 0.05}  &  {\it 16.38 $\pm$ 0.05}  &  {\it 15.47 $\pm$ 0.05} & {\it 15.73 $\pm$ 0.06}  & {\it 16.02 $\pm$ 0.07}  & {\it 15.94 $\pm$ 0.06}  \\
2013-01-15  &  56307  &  {\it 15.73 $\pm$ 0.05}  &  {\it 16.32 $\pm$ 0.05}  &  {\it 15.46 $\pm$ 0.05} & {\it 15.72 $\pm$ 0.06}  & {\it 16.00 $\pm$ 0.06}  & {\it 15.93 $\pm$ 0.06}  \\
2013-02-15  &  56338  &  {\it 15.81 $\pm$ 0.05}  &  {\it 16.33 $\pm$ 0.05}  &  {\it 15.51 $\pm$ 0.05} & {\it 15.78 $\pm$ 0.06}  & {\it 16.14 $\pm$ 0.07}  & {\it 16.07 $\pm$ 0.06}  \\
2013-02-15  &  56338  &  {\it 15.82 $\pm$ 0.05}  &  {\it 16.38 $\pm$ 0.05}  &  {\it 15.55 $\pm$ 0.05} & {\it 15.83 $\pm$ 0.06}  & {\it 16.20 $\pm$ 0.07}  & {\it 16.09 $\pm$ 0.06}  \\
\hline
\end{tabular}
\end{center}
\end{table*} 

\addtocounter{table}{-2}
\begin{table*}
 \begin{center}
\begin{tabular}{cccccccc}
\hline
\multicolumn{1}{c}{\textbf{Date (UT)}} &
\multicolumn{1}{c}{\textbf{MJD}}       &
\multicolumn{1}{c}{\textbf{$v$}}       &
\multicolumn{1}{c}{\textbf{$b$}}       &
\multicolumn{1}{c}{\textbf{$u$}}       &
\multicolumn{1}{c}{\textbf{$w1$}}      &
\multicolumn{1}{c}{\textbf{$m2$}}      &
\multicolumn{1}{c}{\textbf{$w2$}}      \\  
\hline 
2018-08-16  &  58346  &           $-$          &          $-$            &         $-$            &        $-$             & {\it 15.94 $\pm$ 0.07}  &           $-$              \\
2018-08-23  &  58353  &           $-$          &          $-$            &         $-$            &        $-$             &          $-$            &   {\it 15.77 $\pm$ 0.06}   \\
2018-08-30  &  58360  &           $-$          &          $-$            &  15.52 $\pm$ 0.05      &        $-$             &          $-$            &           $-$              \\
2018-09-06  &  58367  &           $-$          &          $-$            &         $-$            & {\it 15.65 $\pm$ 0.06} &          $-$            &           $-$              \\
2018-09-13  &  58374  &           $-$          &          $-$            &         $-$            &        $-$             &    15.94 $\pm$ 0.07     &           $-$              \\
2018-09-20  &  58381  &           $-$          &          $-$            &         $-$            &        $-$             &          $-$            &   15.89 $\pm$ 0.06         \\
2018-09-27  &  58388  &           $-$          &          $-$            & {\it 15.58 $\pm$ 0.05} &        $-$             &          $-$            &           $-$              \\
\hline
\end{tabular}
\end{center}
\end{table*} 

\setcounter{table}{1}
\begin{table*}
\caption{Observed magnitude of SBS\,0846$+$513 obtained by {\em Swift}--UVOT.}
\label{UVOT_0846}
\begin{center}
\begin{tabular}{cccccccc}
\hline
\multicolumn{1}{c}{\textbf{Date (UT)}} &
\multicolumn{1}{c}{\textbf{MJD}}       &
\multicolumn{1}{c}{\textbf{$v$}}       &
\multicolumn{1}{c}{\textbf{$b$}}       &
\multicolumn{1}{c}{\textbf{$u$}}       &
\multicolumn{1}{c}{\textbf{$w1$}}      &
\multicolumn{1}{c}{\textbf{$m2$}}      &
\multicolumn{1}{c}{\textbf{$w2$}}      \\  
\hline
2011-08-30 & 55803 &       $-$               &          $-$           &        $-$             & {\it 19.32 $\pm$ 0.08} &        $-$              &           $-$            \\  
2011-09-15 & 55819 &  {\it 18.78 $\pm$ 0.23} & {\it 19.67 $\pm$ 0.20} & {\it 19.09 $\pm$ 0.16} & {\it 19.21 $\pm$ 0.15} & {\it 19.39 $\pm$ 0.16}  & {\it 19.35 $\pm$ 0.12}   \\
2011-12-20 & 55915 &  $>$ 18.64              &  $>$ 19.81             &  $>$ 19.44             & {\it 19.86 $\pm$ 0.22} & 19.53 $\pm$ 0.39        & 19.84 $\pm$ 0.29         \\  
2011-12-27 & 55922 &       $-$               &  20.13 $\pm$ 0.42      &  19.85 $\pm$ 0.43      & 19.88 $\pm$ 0.35       & 19.58 $\pm$ 0.23        & 20.06 $\pm$ 0.31         \\  
2011-12-28 & 55923 &  $>$ 18.91              &  $>$ 19.92             &  19.12 $\pm$ 0.31      & {\it 19.81 $\pm$ 0.16} & 19.84 $\pm$ 0.40        & 20.11 $\pm$ 0.34         \\  
2012-01-03 & 55929 &       $-$               &  $-$                   &         $-$            &         $-$            &        $-$              & 20.32 $\pm$ 0.39         \\  
2012-05-27 & 56074 &  {\it 16.51 $\pm$ 0.08} & {\it 17.22 $\pm$ 0.07} & {\it 16.79 $\pm$ 0.08} & {\it 17.35 $\pm$ 0.10} & {\it 17.28 $\pm$ 0.10}  & {\it 17.58 $\pm$ 0.09}   \\
\hline
\end{tabular}
\end{center}
\end{table*}

\setcounter{table}{2}
\begin{table*}
\caption{Observed magnitude of PMN\,J0948$+$0022 obtained by {\em Swift}--UVOT.}
\label{UVOT_0948}
\begin{center}
\begin{tabular}{cccccccc}
\hline
\multicolumn{1}{c}{\textbf{Date (UT)}} &
\multicolumn{1}{c}{\textbf{MJD}}       &
\multicolumn{1}{c}{\textbf{$v$}}       &
\multicolumn{1}{c}{\textbf{$b$}}       &
\multicolumn{1}{c}{\textbf{$u$}}       &
\multicolumn{1}{c}{\textbf{$w1$}}      &
\multicolumn{1}{c}{\textbf{$m2$}}      &
\multicolumn{1}{c}{\textbf{$w2$}}      \\  
\hline
2008-12-05 &  54805  &  {\it 18.08 $\pm$ 0.13} &  {\it 18.51 $\pm$ 0.10} &  {\it 17.72 $\pm$ 0.08} &  {\it 17.48 $\pm$ 0.06} &   {\it 17.64 $\pm$ 0.09} &  {\it 17.62 $\pm$ 0.11}  \\ 
2009-03-26 &  54916  &  {\it 17.56 $\pm$ 0.07} &  {\it 17.90 $\pm$ 0.07} &  {\it 17.05 $\pm$ 0.06} &  {\it 17.06 $\pm$ 0.07} &   {\it 17.05 $\pm$ 0.08} &  {\it 17.05 $\pm$ 0.06}  \\
2009-04-15 &  54936  &  {\it 17.82 $\pm$ 0.11} &  {\it 18.22 $\pm$ 0.08} &  {\it 17.19 $\pm$ 0.07} &  {\it 17.07 $\pm$ 0.07} &   {\it 17.02 $\pm$ 0.08} &  {\it 17.13 $\pm$ 0.10}  \\ 
2009-05-05 &  54956  &  {\it 17.23 $\pm$ 0.07} &  {\it 17.74 $\pm$ 0.07} &  {\it 16.80 $\pm$ 0.06} &  {\it 16.86 $\pm$ 0.07} &   {\it 16.94 $\pm$ 0.07} &  {\it 16.95 $\pm$ 0.06}  \\ 
2009-05-10 &  54961  &  {\it 17.87 $\pm$ 0.11} &  {\it 18.13 $\pm$ 0.08} &  {\it 17.23 $\pm$ 0.07} &  {\it 17.10 $\pm$ 0.07} &   {\it 17.18 $\pm$ 0.08} &  {\it 17.19 $\pm$ 0.16}  \\ 
2009-05-15 &  54966  &  {\it 18.26 $\pm$ 0.26} &  {\it 18.31 $\pm$ 0.16} &  {\it 17.62 $\pm$ 0.14} &  {\it 17.34 $\pm$ 0.10} &   {\it 17.50 $\pm$ 0.13} &  {\it 17.64 $\pm$ 0.10}  \\ 
2009-05-25 &  54976  &  {\it 17.78 $\pm$ 0.12} &  {\it 18.29 $\pm$ 0.09} &  {\it 17.37 $\pm$ 0.07} &  {\it 17.21 $\pm$ 0.07} &   {\it 17.20 $\pm$ 0.08} &  {\it 17.29 $\pm$ 0.06}  \\
\hline
\end{tabular}
\end{center}
\end{table*}  

\setcounter{table}{3}
\begin{table*}
\caption{Observed magnitude of PKS\,1502$+$036 obtained by {\em Swift}--UVOT.}
\label{UVOT_1502}
\begin{center}
\begin{tabular}{cccccccc}
\hline
\multicolumn{1}{c}{\textbf{Date (UT)}} &
\multicolumn{1}{c}{\textbf{MJD}}       &
\multicolumn{1}{c}{\textbf{$v$}}       &
\multicolumn{1}{c}{\textbf{$b$}}       &
\multicolumn{1}{c}{\textbf{$u$}}       &
\multicolumn{1}{c}{\textbf{$w1$}}      &
\multicolumn{1}{c}{\textbf{$m2$}}      &
\multicolumn{1}{c}{\textbf{$w2$}}      \\  
\hline
2009-07-25  & 55037 & {\it 18.98 $\pm$ 0.22} & {\it 19.47 $\pm$ 0.17} & {\it 18.60 $\pm$ 0.13} & {\it 18.46 $\pm$ 0.11}   & {\it 18.37 $\pm$ 0.11}     & {\it 18.38 $\pm$ 0.08} \\ 
2012-04-25  & 56042 & 18.62 $\pm$ 0.35       & 19.76 $\pm$ 0.40       & 19.11 $\pm$ 0.33       & 18.89 $\pm$ 0.24         & {\it 18.81 $\pm$ 0.09}     & 18.49 $\pm$ 0.15       \\ 
2012-05-25  & 56072 & 18.67 $\pm$ 0.35       & 19.59 $\pm$ 0.32       & {\it 18.59 $\pm$ 0.06} & 18.22 $\pm$ 0.16         & 18.45 $\pm$ 0.19           & 18.18 $\pm$ 0.12       \\ 
2012-06-25  & 56103 & 18.74 $\pm$ 0.39       & 19.23 $\pm$ 0.33       & 18.72 $\pm$ 0.31       & {\it 18.32 $\pm$ 0.07}   & 18.16 $\pm$ 0.18           & 18.51 $\pm$ 0.16       \\ 
2012-08-07  & 56146 & {\it 19.46 $\pm$ 0.37} & {\it 20.09 $\pm$ 0.30} & {\it 18.95 $\pm$ 0.18} & {\it 18.61 $\pm$ 0.13}   & {\it 18.51 $\pm$ 0.12}     & {\it 18.50 $\pm$ 0.09} \\
2012-08-08  & 56147 &  $>$ 18.81             & 19.15 $\pm$ 0.33       & 18.82 $\pm$ 0.36       & 19.12 $\pm$ 0.34         & 18.40 $\pm$ 0.20           & 18.45 $\pm$ 0.15       \\ 
2015-12-22  & 57378 & 18.26 $\pm$ 0.22       & 18.52 $\pm$ 0.18       & {\it 17.84 $\pm$ 0.11} & {\it 17.72 $\pm$ 0.10}   & 17.79 $\pm$ 0.10           & 17.93 $\pm$ 0.11       \\
\hline
\end{tabular}
\end{center}
\end{table*}

\setcounter{table}{4}
\begin{table*}
\caption{Observed magnitude of FBQS\,J1644$+$2619 obtained by {\em Swift}--UVOT.}
\label{UVOT_1644}
\begin{center}
\begin{tabular}{cccccccc}
\hline
\multicolumn{1}{c}{\textbf{Date (UT)}} &
\multicolumn{1}{c}{\textbf{MJD}}       &
\multicolumn{1}{c}{\textbf{$v$}}       &
\multicolumn{1}{c}{\textbf{$b$}}       &
\multicolumn{1}{c}{\textbf{$u$}}       &
\multicolumn{1}{c}{\textbf{$w1$}}      &
\multicolumn{1}{c}{\textbf{$m2$}}      &
\multicolumn{1}{c}{\textbf{$w2$}}      \\
\hline
2011-12-26 & 55921  & 17.64 $\pm$ 0.22       & 18.18 $\pm$ 0.16       & 16.80 $\pm$ 0.10       & 16.97 $\pm$ 0.11       & 16.88 $\pm$ 0.12       & {\it 16.92 $\pm$ 0.07}  \\  
2015-04-08 & 57120  &         $-$            &    $-$                 & 17.32 $\pm$ 0.07       &        $-$             &      $-$               &        $-$              \\ 
2015-04-09 & 57121  & {\it 17.44 $\pm$ 0.16} & {\it 18.02 $\pm$ 0.13} & {\it 17.31 $\pm$ 0.12} & {\it 17.23 $\pm$ 0.13} & {\it 17.35 $\pm$ 0.13} & {\it 17.26 $\pm$ 0.07}  \\ 
2015-05-05 & 57147  & 17.88 $\pm$ 0.28       & 18.26 $\pm$ 0.19       & 17.24 $\pm$ 0.15       & {\it 17.45 $\pm$ 0.08} & 17.52 $\pm$ 0.18       & 17.49 $\pm$ 0.14        \\ 
2015-06-05 & 57178  &        $-$             &    $-$                 &      $-$               & {\it 17.36 $\pm$ 0.08} & 17.52 $\pm$ 0.12       &      $-$                \\ 
2015-07-05 & 57208  &        $-$             &    $-$                 & {\it 17.17 $\pm$ 0.05} &     $-$                &      $-$               &      $-$                \\ 
2015-08-05 & 57239  & {\it 17.55 $\pm$ 0.15} & {\it 17.95 $\pm$ 0.12} & {\it 16.90 $\pm$ 0.09} & {\it 17.04 $\pm$ 0.07} & {\it 17.07 $\pm$ 0.11} & {\it 16.98 $\pm$ 0.09}  \\
\hline
\end{tabular}
\end{center}
\end{table*}

\setcounter{table}{5}
\begin{table*}
\caption{Observed magnitude of PKS\,2004--447 obtained by {\em Swift}--UVOT.}
\label{UVOT_2004}
\begin{center}
\begin{tabular}{cccccccc}
\hline
\multicolumn{1}{c}{\textbf{Date (UT)}} &
\multicolumn{1}{c}{\textbf{MJD}}       &
\multicolumn{1}{c}{\textbf{$v$}}       &
\multicolumn{1}{c}{\textbf{$b$}}       &
\multicolumn{1}{c}{\textbf{$u$}}       &
\multicolumn{1}{c}{\textbf{$w1$}}      &
\multicolumn{1}{c}{\textbf{$m2$}}      &
\multicolumn{1}{c}{\textbf{$w2$}}      \\  
\hline
2011-05-15 &   55696  &  {\it 18.71 $\pm$ 0.11}  &  {\it 19.93 $\pm$ 0.19}  &  {\it 19.08 $\pm$ 0.13}  &  {\it 19.86 $\pm$ 0.21} &  {\it 20.26 $\pm$ 0.28} &  {\it 20.49 $\pm$ 0.26}  \\ 
2011-07-14 &   55756  &  {\it 19.35 $\pm$ 0.21}  &  {\it 19.83 $\pm$ 0.16}  &  {\it 19.27 $\pm$ 0.13}  &  {\it 19.92 $\pm$ 0.24} &  {\it 20.17 $\pm$ 0.30} &  {\it 20.26 $\pm$ 0.26}  \\ 
2011-07-19 &   55761  &  18.67 $\pm$ 0.31        &         $-$              &  $>$ 18.96               &  $>$ 19.40              &  $>$ 19.29              &  $>$ 19.63               \\ 
2011-07-23 &   55765  &  19.63 $\pm$ 0.45        &  20.58 $\pm$ 0.47        &  19.27 $\pm$ 0.22        &  19.60 $\pm$ 0.32       &  $>$ 20.18              &  19.84 $\pm$ 0.33        \\ 
2011-08-10 &   55783  &     $-$                  &        $-$               &      $-$                 &  19.57 $\pm$ 0.24       &  $-$                    &         $-$              \\ 
2011-08-13 &   55786  &  18.69 $\pm$ 0.33        &  19.56 $\pm$ 0.31        &  19.32 $\pm$ 0.34        &  19.68 $\pm$ 0.37       &  {\it 20.05 $\pm$ 0.26} &  20.31 $\pm$ 0.35        \\ 
2011-09-05 &   55809  &  18.64 $\pm$ 0.47        &  19.58 $\pm$ 0.51        &  18.85 $\pm$ 0.40        &  19.20 $\pm$ 0.42       &  $>$ 19.55              &  {\it 20.17 $\pm$ 0.30}  \\
\hline
\end{tabular}
\end{center}
\end{table*}

The average magnitudes in each UVOT filter for all sources are reported in Table~\ref{Average_summed}. We have also calculated the long-term variability amplitude (V$_{\rm\,amp,\,LT}$) for each filter, where V$_{\rm\,amp,\,LT}$ is calculated as the ratio of maximum to minimum flux as $F_{\rm\,max}/F_{\rm\,min}$ = 2.581164$^{\Delta\,mag_{\rm\,max}}$, where $\Delta\,mag_{\rm\,max}$ = mag$_{\rm\,max}$ -- mag$_{\rm\,min}$, and mag$_{\rm\,max}$ and mag$_{\rm\,min}$ are the maximum and minimum magnitude observed, respectively, over the entire {\em Swift} monitoring period. 

\begin{table}
\caption{Average magnitude of the $\gamma$-ray-emitting NLSy1 in the UVOT filters based on summed images obtained by summing the single exposures collected at each epoch.}
\centering
\begin{tabular}{ccccccc}
\hline
\textbf{Source name} & \textbf{$v$} & \textbf{$b$} &\textbf{$u$} & \textbf{$w1$} & \textbf{$m2$} & \textbf{$w2$} \\
\hline
1H\,0323$+$342     &  15.65  & 16.19 & 15.34 & 15.56 & 15.89 & 15.78 \\
SBS\,0846$+$513    &  18.58  & 19.48 & 19.07 & 19.24 & 19.09 & 19.33 \\
PMN\,J0948$+$0022  &  17.76  & 18.12 & 17.32 & 17.13 & 17.20 & 17.21 \\
PKS\,1502$+$036    &  18.76  & 19.30 & 18.64 & 18.37 & 18.32 & 18.32 \\
FBQS\,J1644$+$2619 &  17.59  & 18.05 & 17.12 & 17.14 & 17.23 & 17.15 \\
PKS\,2004--447     &  18.88  & 19.70 & 18.97 & 19.54 & 19.84 & 20.05 \\
\hline
\end{tabular}
\label{Average_summed} 
\end{table} 
 
\begin{table}
\caption{Long-term variability amplitude of the $\gamma$-ray-emitting NLSy1 in the UVOT filters based on summed images obtained by summing the single exposures collected at each epoch.} 
\centering 
\begin{tabular}{ccccccc}
\hline
\textbf{Source name} & \textbf{$v$} & \textbf{$b$} &\textbf{$u$} & \textbf{$w1$} & \textbf{$m2$} & \textbf{$w2$} \\ 
\hline 
1H\,0323$+$342     & 1.45  &  1.64  &  1.89  &  1.96  & 2.12  &  2.16  \\ 
SBS\,0846$+$513    & 18.36 & 23.12  & 24.21  & 21.68  & 21.88 &  20.32 \\ 
PMN\,J0948$+$0022  & 10.96 &  5.86  &  5.97  &  3.31  & 5.25  &  4.45  \\ 
PKS\,1502$+$036    &  3.02 &  4.25  &  3.60  &  3.63  & 2.56  &  1.75  \\ 
FBQS\,J1644$+$2619 &  1.61 &  1.38  &  1.61  &  1.63  & 1.80  &  1.69  \\ 
PKS\,2004--447     &  3.37 &  4.29  &  2.49  &  2.61  & 2.91  &  2.54  \\ 
\hline 
\end{tabular} 
\label{Amplitude_summed}  
\end{table}  

\clearpage

\begin{figure*}
\begin{center}
{\includegraphics[width=0.75\textwidth]{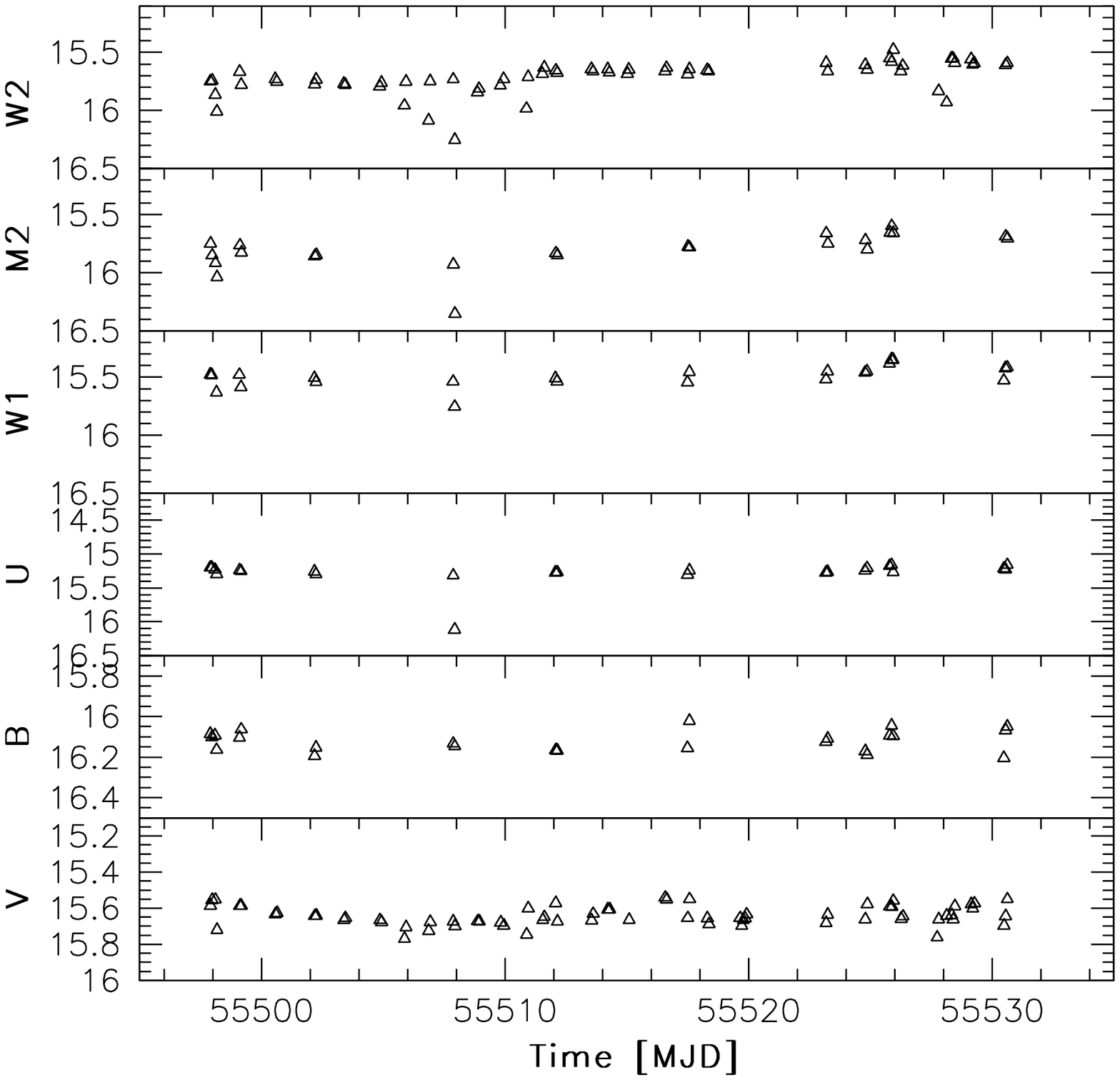}}
\caption{Light curve of 1H\,0323$+$342 in the UVOT filters for the period 2010 October 28 (MJD 55497)--November 30 (MJD 55530) considering single exposures. Errors are quite small (0.04--0.06 mag) and are not shown in the plot.}
\label{0323_lc}
\end{center}
\end{figure*}

\end{document}